\documentclass[runningheads,a4paper]{llncs}

\usepackage{booktabs} 
\usepackage{latexsym}
\usepackage{amsmath,amsfonts,amssymb}
\usepackage{stmaryrd}
\usepackage{wasysym}
\usepackage{graphicx}
\usepackage{extarrows}
\usepackage{enumitem}
\usepackage{algorithm}
\usepackage{algorithmic}
\usepackage{multirow}
\usepackage{xcolor}
\usepackage{float}
\usepackage{wrapfig,lipsum,booktabs}
\usepackage{times}
\usepackage{bm}
\usepackage{listings}
\usepackage{xcolor}
\usepackage{subfig}

\usepackage{amsmath}
\usepackage{stmaryrd}
\usepackage{mathbbol}
\usepackage{yhmath}
\usepackage{url}
\usepackage{times}
\usepackage{latexsym,amsmath,epsfig,amssymb,graphics}
\usepackage{color,graphicx}
\usepackage{amsfonts}
\usepackage{graphpap}
\usepackage{listings}

\allowdisplaybreaks[4]

\newcommand{\pskip}{\textmd{skip}}
\newcommand{\nstop}{\textbf{stop}}

\newcommand{\pstop}{\epsilon}

\newcommand{\evolutiondde}[3]{\langle {#1}(\dot{#2}(t),#2(t), #2(t-r_1), ...,#2(t-r_k))=0 \& #3\rangle}
\newcommand{\evolutionddesem}[3]{\langle {#1}(\dot{#2}(t),...,#2(t-r_k))=0 \& #3\rangle}

\newcommand{\evolutionn}[2]{\langle #1 \& #2\rangle}

\newcommand{\pwait}{\textrm{wait}}
\newcommand{\exempt}[4]{#1 \unrhd \talloblong_{#2} (#3 \rightarrow #4)}

\newcommand{\discrete}[3]{\textit{D}_{#1, #2}(#3)}

\newcommand{\compress}[3]{#1\stackrel{#2}{\twoheadrightarrow}#3}
\newcommand{\fracN}[2]{\frac{\small \textstyle #1}{\small \textstyle #2}}
\newcommand{\bisimilar}[2]{#1\cong_{h,\varepsilon}#2}
\newcommand{\euler}{\xi}

\def \dhcsp {{\emph{d}HCSP}}

\newcommand{\RR}{\mathbb{R}}

\newcommand{\s}{\mathbf{s}}

\newcommand{\yy}{\mathbf{y}}

\newcommand{\xx}{\mathbf{x}}

\newcommand{\ttt}{\mathbf{t}}
\newcommand{\dd}{\mathbf{d}}

\newcommand{\ee}{\mathbf{e}}

\newcommand{\ff}{\mathbf{f}}

\newcommand{\fg}{\mathbf{g}}

\newcommand{\vare}{\varepsilon}

\newcommand{\Define}{\stackrel{\mbox{\small\rm def}}{=}}

\newcommand{\repnum}{\textit{num}}

\newcommand{\oomit}[1]{}
 \newcommand{\Intv}{\textit{Intv}}
\input{Userdef.sty}

\newcommand{\variable}{\textit{Var}}

\newcommand{\staten}{\rho}

\usepackage{url}
\urldef{\mailsa}\path|{yangg, ljiao, wangsl, znj}@ios.ac.cn|
\newcommand{\keywords}[1]{\par\addvspace\baselineskip
\noindent\keywordname\enspace\ignorespaces#1}
\oomit{\urldef{\mailsa}\path|{yangg,ljiao,wanglt,wangsl,znj}@ios.ac.cn|}

\graphicspath{{figures/}}

\begin{document}

\title{Synthesizing SystemC Code from Delay Hybrid CSP %
\thanks{This work is partially supported by ``973 Program" under grant No. 2014CB340701, by NSFC under grants 61625206, 61732001 and 91418204, by CDZ project CAP (GZ 1023), and by the CAS/SAFEA International Partnership Program for Creative Research Teams.}%
}
\titlerunning{Generating Reliable SystemC Code from Time-Delayed Hybrid CSP}
\author{Gaogao Yan \textsuperscript{1,2} \and Li Jiao \textsuperscript{1} \and Shuling  Wang \textsuperscript{1} \and Naijun Zhan \textsuperscript{1,2}}
\authorrunning{Gaogao Yan , Li Jiao, Shuling  Wang, and Naijun Zhan}
\institute{\textsuperscript{1}State Key Lab. of Comput. Sci.,
Institute of Software, Chinese Academy of Sciences \\
\textsuperscript{2}University of Chinese Academy of Sciences \\
\mailsa\\}

\toctitle{Lecture Notes in Computer Science}
\tocauthor{Authors' Instructions}
\maketitle
\pagestyle{empty}

\begin{abstract}
%
 \oomit{ we first extend the discretization of HCSP processes with the globally asymptotic stability (GAS) condition proposed in~\cite{Yan16} to general HCSP processes without GAS, but w.r.t. bounded time, and prove that the discretized HCSP model and the original HCSP model are approximately bisimilar.   Based on the discretization, we then define a set of rules to transform a discrete HCSP model to a piece of SystemC code such that they are bisimilar. We  implement a tool to automatically do the translation  from HCSP processes to SystemC code. The approach is illustrated by some real-world case studies.}

Delay is omnipresent  in modern control systems, which can
  prompt oscillations
  and may
  cause
  deterioration of control performance, invalidate both
  stability and safety properties.
 This implies that safety or stability certificates
obtained on idealized, delay-free models of systems prone to delayed
coupling may be erratic, and further the incorrectness of
the executable code generated from these models. However, automated methods for system verification
 and code generation that
ought to address models of system dynamics reflecting delays have not been paid enough attention yet
in the computer science community.
In our previous work, on one hand, we investigated the verification of delay dynamical and hybrid systems; on the
other hand, we also addressed how to synthesize SystemC code from a verified hybrid system modelled by Hybrid CSP (HCSP)
without delay. In this paper, we give a first attempt to
synthesize SystemC code from a verified delay hybrid system modelled by Delay HCSP ({\dhcsp}), which is an extension
of HCSP by replacing ordinary differential equations (ODEs) with delay differential equations (DDEs).
\oomit{ In model-based engineering, how to generate reliable code from abstract control models, that are usually described by hybrid systems, is challenging, as  hybrid systems involve continuous and discrete dynamics, and their interactions, while code only contains discrete actions. In our previous work, we present a discretization of hybrid systems modelled by hybrid CSP (HCSP), an extension of CSP by introducing ordinary differential equations (ODEs) for specifying continuous evolution,  and then based on  approximate bisimulation, prove the approximate  equivalence between the source and target models within given precision. In this paper, we extend this work, to consider hybrid systems with time-delayed dynamics, that are specified by delayed differential equations (DDEs) instead of ODEs. The main difference comes from the approximation of the trajectories of DDEs by  discrete states  and the validated computation of the error bounds of the discretization. Based on existing algorithms we have developed, we present in this paper a discretization of delayed HCSP (\dhcsp), the extension of HCSP with DDEs, and furthermore, translate discretized \dhcsp to reliable SystemC code at implementation level.}
  We implement a tool to support the automatic translation from {\dhcsp} to SystemC.

\end{abstract}

\section{Introduction}\label{sec:introduction}
Model-Driven Design (MDD) is considered as an effective way of developing reliable complex
 embedded systems (ESs), and has been successfully applied in industry \cite{HS06,Lee00}, therefore drawn
 increasing attentions recently. A challenging problem in MDD  is
  to transform a verified abstract model at high-level step by step to more concrete  models at lower levels, and to executable code at the end.  To make sure that the final code generated in MDD is correct and
  reliable, the transformation process must be guaranteed to preserve consistency between observational behaviors of the models at different levels in a rigorous way. However, this is difficult, due to the inherent complexity of most ESs, especially for hybrid systems, which contain complicated behaviour, like both continuous and discrete dynamics, and the complex interactions between them, time-delay, and so on, while code only contains discrete actions. Obviously, the exact equivalence between them can never be achieved, due to the unavoidable error of discretization of continuous dynamics of hybrid systems.

As an effective way for analyzing  hybrid systems and their discretization,
approximate bisimulation~\cite{Girard07a} can solve the above problem.
Instead of requiring observational behaviors of two systems
to be exactly identical, it allows errors but requires the ¡°distance¡± between two systems
remains bounded by some precisions. In our pervious work~\cite{Yan16}, we used Hybrid CSP (HCSP), an extension of CSP by introducing
differential equations (DEs) for modelling continuous evolutions and
interrupts for modelling interaction between continuous and discrete
dynamics, as the modelling language for hybrid systems; and then,  we
extended the notion of approximate bisimulation to general hybrid
systems modelled as HCSP processes;  lastly, we
presented an algorithm to discretize an HCSP process (a control
model) by a discrete HCSP process (an algorithm model), and
proved that they are approximately bisimilar if the original HCSP
process satisfies the   globally asymptotical stability (GAS) condition. Here the
GAS condition requires the DEs starting from any initial state
can always infinitely approach to its equilibrium point as time
proceeds~\cite{TAC02Angeli}.  Recently, in \cite{Yan17}, we further considered how to
discretize  an HCSP process without GAS, and refine the discretized HCSP process to
SystemC code, which is approximately bisimilar to the original HCSP process in a given bounded time.

On the other hand, in practice, delay is omnipresent  in modern control systems.
For instance, in a distributed real-time control system, control commands may depend on communication with sensors and actuators over a communication network introducing latency.
\oomit{Such setups of control have become standard in, e.g., several applications of industrial robots \cite{low2007industrial}. Beyond networks connecting and thereby delaying signals from physical sensors, the subsequent digital signal processing also increasingly adds to delays due to its ever-growing complexity.
It is intuitively obvious that delayed observation renders control harder: the controller has to decide based on dated information, which may no longer be fully indicative of the current situation, thus possibly issuing an inadequate reaction or keeping up an initially adequate one for too long, thereby inducing oscillations and similar phenomena.}
 This implies that safety or stability certificates
obtained on idealized, delay-free models of systems prone to delayed
coupling may be erratic, and further the incorrectness of
the code generated from these models. However, automated methods for system verification
 and code generation that
ought to address models of system dynamics reflecting delays have not been paid enough attention yet
in the computer science community.

Zou \textit{et al}. proposed in \cite{Zou15} a safe enclosure method to automatic stability analysis and
 verification of delay differential equations by using interval-based
Taylor over-approximation to enclose a set of functions by a parametric Taylor series
with parameters in interval form. Prajna \textit{et al}. extended
the barrier certificate method for ODEs to the polynomial time-delay differential
equations setting, in which the safety verification problem is formulated as a problem of
solving sum-of-square programs \cite{PJ05}. Huang \textit{et al}. presents a technique for simulation based
time-bounded invariant verification of nonlinear networked dynamical systems
with delayed interconnections by computing bounds on the sensitivity of trajectories
(or solutions) to changes in initial states and inputs of the system \cite{Huang17}. A similar simulation
method integrating error analysis of the numeric solving and the sensitivity-related state
bloating algorithms was proposed in \cite{Chen16} to obtain safe enclosures of time-bounded
reach sets for systems modelled by DDEs.

\oomit{
However, there are three problems not addressed by~\cite{Yan16}: first, the GAS condition on DEs is sometimes too restrictive;
second, the discrete HCSP is not executable code; the last but not the least, only the general  ordinary differential equations (ODEs) are considered in the discretization. In practice, there are a broad class of time-delayed dynamic systems that cannot be specified by ODEs. The character of time-delayed dynamics is that, the future evolution of the dynamics is not only
governed by current state, but depends on its past states. Such dynamic systems are very common, e.g., for the population dynamics, the birth rate follow changes in population size with a delay to reproductive age; for network control systems, the delay is introduced because of the transport delay when transmitting data in the communication network, and so on. Instead of ODEs, delayed differential equations (DDEs) are proposed for characterizing such dynamics~\cite{Bellman1963}.

In order to address the problem of generating correct-by-construction controllers for control applications entailing such delay in sensing, computation of control reaction, or actuation, we investigate strategies in safety games that are \emph{winning when played under delay}. In this setting the controller is enforced to make decisions before being aware of the entire history of a play. Hence, playing the game under $\delta$ delay means that the winning strategy has to pre-decide control actions at a predecessor state $\delta$ time units before the action shall take effect. We show in this paper, that the existence of a winning strategy for the controller in games with delays is decidable with respect to safety objectives, demonstrate that in contrast to undelayed safety games it cannot be memoryless in general, and derive the necessary amount of memory (which is uniform and finite). Aiming at addressing the controller synthesis problem, we furthermore present an algorithm to successively synthesize a series of controllers over increasing delays, starting from the most permissive controller for the undelayed safety game. Due to monotonicity properties, such incremental synthesis seems favorable to reduction of the problem to an exponentially larger in the delay safety game, as incremental synthesis successively strips the state space from all parts not controllable under a given --and thus any larger-- delay. Moreover, we obtain an upper bound on the delay, such that the existence of a corresponding winning strategy implies controllability of the game under any arbitrarily large delay. \oomit{We believe that our approach will pave the way for synthesizing efficient controllers in safety critical applications, where the controllers are exposed to considerable delays. Also, an extension for hybrid control is under development and will be exposed in future work.}}

However, in the literature, there is few work on how to refine a verified ES model with delay to executable code in MDD.
In this paper, we address this issue, and the main contributions can be summarized as follows:
\begin{itemize}
\item First of all, we extend HCSP by allowing delay, called Delay HCSP ({\dhcsp}),
which is achieved by replacing ODEs with DDEs in HCSP. Obviously, HCSP is a proper subset of
{\dhcsp} as all ODEs can be seen as specific DDEs in which time delay is zero.
Then, we propose the notion of \emph{approximately bisimilar} over {\dhcsp} processes.

\oomit{\item Instead of the GAS requirement on DEs, we
consider a hybrid system executed within a bounded time, without any restriction on DEs.
By dropping the GAS condition, we show that whether
two {\dhcsp} processes are approximately bisimilar within
bounded time is decidable.}

\item  In ~\cite{Chen16}, the authors presented an approach to discretizing a DDE by a sequence of states corresponding to discrete time-stamps and meanwhile the error bound that defines the distance from the trajectory is computed automatically on-the-fly. As a result, by adjusting step size of the discretization, the given precision can be guaranteed.  Inspired by
     their work, we consider how to discretize a {\dhcsp} process $S$ such that
    the discretized {\dhcsp} process is approximately bisimilar to $S$.
    This is done
     by defining a set of rules and proving that any {\dhcsp} process $S$ and its discretization are approximately bisimilar within  bounded time with respect to the given precision.

\item Finally, we present a set of code generation rules from
discrete {\dhcsp}  to executable SystemC code and prove the equivalence between them.
\end{itemize}

We implement a prototypical tool to automatically transform  a {\dhcsp} process to SystemC code
and provide some case studies to illustrate the above approach. Due to space limitation, the proofs of theorems are available in Appendix A.

%
%
%
%
%

\subsection{Related work}
Generating reliable code from control models is a dream of embedded engineering but difficult. For some popular models such as Esterel \cite{Esterel}, Statecharts \cite{Statecharts}, and Lustre \cite{Lustre1}, code generation is supported. However, they do not take continuous behavior into consideration. Code generation is also supported in some commercial tools such as Simulink \cite{slusing}, Rational Rose \cite{RationalRose}, and TargetLink \cite{TargetLink}, but the correctness between the model and the code generated from it is not formally guaranteed, as they mainly focus on the numerical errors. The same issue exists in SHIFT \cite{SHIFT}, a modelling language for hybrid automata. Generating code from a special hybrid model, CHARON \cite{CHARON}, was studied in \cite{singlethread,multithread,TOC2010}. Particularly, in order to ensure the correctness between a CHARON model and its generated code, a formal criteria \emph{faithful implementation} is proposed in \cite{TOC2010}, but it can only guarantee the code model is under-approximate to the original hybrid model. The main difference between the above works and ours lies in that
the delayed dynamics is considered for the code generation from hybrid models in our work.

For the discretization of DDEs, we can refer to some existing works which focus on the verification of systems containing
 delayed differential dynamics.
 In \cite{Zou15}, a method for analyzing the stability and safety of a special class of DDEs was proposed,
  which cannot deal with the mixed ODE-DDE form. In \cite{Pola10}, the authors proposed a method for constructing a symbolic model from an incrementally input-to-state stable ($\delta$-ISS) nonlinear time-delay system, and moreover proved the symbolic model and the original model are approximately bisimilar. After that, they proved the same result for the incrementally input-delay-to-state stable ($\delta$-IDSS) nonlinear time-delay system with unknown and time-varying delays in \cite{Pola15}. Unfortunately, the $\delta$-ISS and $\delta$-IDSS condition are difficult to check in practice. A simulation-based method is proposed in \cite{Huang17} for computing an over-approximate reachable set of a time-delayed nonlinear networked dynamical system. Within this approach, a significant function (i.e., the IS discrepancy function), used for bounding the distance between two trajectories, is difficult to find for general dynamical systems. In \cite{Chen16}, a further extension of \cite{Huang17} that can handle any kind of DDEs with constant time delays is introduced, which can be appropriately used for the discretization of DDEs in {\dhcsp}. But no work is available on how to generate executable code from a verified model with delay.


The rest of this paper is organized as: Some preliminary
notions on DDEs \oomit{transition systems,  {\dhcsp}}and SystemC are introduced in Sec. \ref{section:Preliminaries}. Sec.  \ref{section:appbisimofhcsp} extends HCSP to {\dhcsp} and defines the approximate
bisimulation on {\dhcsp}.  In Sec. \ref{section:discretizationofhcsp}, the discretization of {\dhcsp} processes is presented and the correctness of the discretization is proved.
  The translation from discrete {\dhcsp} to SystemC code is presented in Sec. \ref{section:codegenerationofhcsp}. In Sec. \ref{section:casestudy}, a case study is provided to illustrate our approach. Sec. \ref{section:conclusion} concludes the paper and discusses the future work.

\section{Preliminaries}
\label{section:Preliminaries}

In this section, we introduce some preliminary knowledge that will be used later.
 \oomit{in this work. Sec. \ref{subsec:DDS} recalls some notions related to delayed dynamical systems, and Sec. \ref{subsec:TS} introduces the transition system based on which the approximate bisimulation relation is defined. The source and target languages, i.e. DHCSP and SystemC, are presented in Sec. \ref{subsec:HCSP} and Sec. \ref{subsec:SystemC}, respectively. }
\subsection{Delay Dynamical Systems}
\label{subsec:DDS}
For a vector $\xx \in \mathbb{R}^n$, $\| \xx\|$ denotes its $L^2$ norm, i.e., $\| \xx \|=\sqrt{x_1^2+x_2^2+...+x_n^2}$.
Given a vector $\xx\in \mathbb{R}^n$ and $\epsilon \in \mathbb{R}^+_0$, $N(\xx, \epsilon)$ is defined as the $\epsilon$-neighbourhood of $\xx$, i.e., $N(\xx, \epsilon)=\{ \yy\in \mathbb{R}^n | \|\xx-\yy\| \le \epsilon\}$. Then,
 for a set $S\subseteq \mathbb{R}^n$, $N(S, \epsilon)$ is defined as $N(S, \epsilon)=\bigcup_{\xx \in S}\{ \yy\in \mathbb{R}^n | \|\xx-\yy\| \le \epsilon\}$, and $conv(S)$ is denoted as the convex hull of $S$. If $S$ is compact, $dia(S)=sup_{\xx,\xx^{\prime} \in S}\|\xx-\xx^{\prime}\|$ defines its diameter.

In this paper, we consider delay dynamical systems governed by the form:
\begin{equation}
\label{eq:dynamical}
\left\{
\begin{array}{ll}
\dot{\xx}(t)=\ff(\xx(t),\xx(t-r_1),...,\xx(t-r_k)),
& t\in [0,\infty ) \\
\xx(t)=\fg(t), & t\in [-r_k,0]
\end{array}\right.
\end{equation}
where $\xx\in \mathbb{R}^n$ is the state, $\dot{\xx}(t)$ denotes the temporal derivative of $\xx$ at time $t$, and $\xx(t)=\fg(t)$ is the \emph{initial condition}, where $\fg$ is assumed to be $\mathrm{C}^0[-r_k,0]$.
 Without loss of generality, we assume the delay terms are ordered as $r_k>...>r_1>0$.

{A function $X(\cdot):[-r_k,\nu)\to \mathbb{R}^n$ is said to be a \emph{trajectory} (solution) of (\ref{eq:dynamical}) on $[-r_k,\nu)$, if $X(t)=\fg(t)$ for all $t \in [-r_k,0]$ and $\dot{X}(t)=\ff(X(t),X(t-r_1),...,X(t-r_k))$ for all $t \in [0,\nu)$.} In order to ensure the existence and uniqueness of the maximal trajectory from a continuous initial condition $\fg(t)$, we assume $\ff$ is continuous and continuously differentiable in the first argument. Then, we write $X(t,\fg(t_0))$ with $t_0\in [-r_k,0]$ to denote the point reached at time $t$ from the initial state $\fg(t_0)$, which should be uniquely determined. Moreover, if $\ff$ is Lipschitz, i.e., there exists a constant $L>0$ s.t. $\| \ff(\xx)-\ff(\yy)\| \le L\| \xx-\yy\|$ holds for all $\xx,\yy$, we can conclude $X(\cdot)$ is unique over $[-r_k,\infty)$. Please refer to \cite{Bellen2013} for the theories of \emph{delay differential equations}.
\subsection{SystemC}
\label{subsec:SystemC}
SystemC is a system-level modelling language supporting both system architecture and software development. It provides a uniform platform for the modelling of complex embedded systems. Essentially it is a set of C++ classes and macros. According to the naming convention of SystemC, most identifiers are prefixed with \emph{SC\_} or \emph{sc\_}, such as   \emph{SC\_THREAD}, \emph{SC\_METHOD}, \emph{sc\_inout}, \emph{sc\_signal}, \emph{sc\_event}, etc.

Modules, denoted by \emph{SC\_MODULE}, are the basic  blocks of a SystemC model. A model usually contains several modules, within which sub-designs, constructors, processes, ports, channels, events and other elements may be included. Each module is defined as a class. The constructor of a module is denoted as \emph{SC\_CTOR()}, in which some initialization operations carry out.
Processes are member functions of the module, describing the actual functionality, and multiple processes execute concurrently in nature.
A process has a list of sensitive events, by whose notifications its execution is controlled.
Two major types of processes, \emph{SC\_METHOD} and \emph{SC\_THREAD}, are supported in SystemC.
Generally, an \emph{SC\_METHOD} can be invoked multiple times, whereas an \emph{SC\_THREAD} can only be invoked once.

\emph{Ports} in SystemC are components using for communicating with each other between modules. They are divided into three kinds by the data direction, i.e., \emph{sc\_in}, \emph{sc\_out} and \emph{sc\_inout} ports. Only ports with the same data type can be connected (via channels). \emph{Channels} are used for connecting different sub-designs, based on which the communication is realized (by calling corresponding methods in channels, i.e., \emph{read()} and \emph{write()}). Channels are declared by \emph{sc\_signal$\langle\rangle$}. Another important element using for synchronization is \emph{event}, which has no value and no duration. Once an event occurs, the processes waiting for it will be resumed. Generally, an event can be notified immediately, one delta-cycle (defined in the execution phase below) later, or some constant time later.
%

The simulation of a SystemC model starts from the entrance of a method named \emph{sc\_main()}, in which three phases are generally involved: elaboration, execution and post-processing.
During the elaboration and the post-processing phase, some initialization and result processing are carried out, respectively. We mainly illustrate the execution phase in the next.

The execution of SystemC models is event-based and it can be divided into four steps: (1) Initialization, executing all concurrent processes in an unspecified order until they are completed or suspended by a \emph{wait()}; (2) Evaluation, running all the processes that are ready in an unspecified order until there are no more ready process; (3) Updating, copying the value of containers (e.g., channels) to the current location, then after that, if any event occurs, go back to step 2. Here, the cycle from evaluation to updating and then go back to evaluation is known as the delta-cycle; (4) Time advancing, if no more processes get ready currently, time advances to the nearest point where some processes will be ready. If no such point exists or the time is greater than a given time bound, the execution will terminate. Otherwise, go back to Step 2.


\section{Delay Hybrid CSP ({\dhcsp}) }
\label{section:appbisimofhcsp}
In this section, we first extend HCSP with delay, and then
discuss the notion of approximate  bisimulation over {\dhcsp} processes by
extending the corresponding notion of HCSP defined in \cite{Yan16}.

\subsection{Syntax of {\dhcsp}}
\label{subsec:HCSP}
{\dhcsp} is an extension of HCSP by introducing DDEs to model continuous evolution with delay behavior. The syntax of {\dhcsp} is given below:
\[
\begin{array}{ll}
  P ::= & \pskip \mid x :=e \mid \pwait \ d \mid   ch?x \mid
         ch!e \mid P;Q  \mid  B \rightarrow P\mid \\
         & P \sqcap Q \mid P^*
 \mid \talloblong_{i\in I} (io_i\rightarrow Q_i)\mid  \evolutiondde{F}{\s}{B} \mid \\
& \exempt{\evolutiondde{F}{\s}{B}}{i\in I}{io_i}{Q_i}\\
  S ::= & P_1 \| P_2 \| \ldots \| P_n \mbox{ for some $n \geq 1$}
  \end{array}
  \]
where $x,\s$ stands for variables and vectors of variables, respectively, $B$ and $e$ are Boolean and arithmetic expressions, $d$ is a non-negative real constant, $ch$ is a channel name, $io_{i}$ stands for a communication event (i.e., either $ch_{i}?x$ or $ch_{i}!e$ for some $x$, $e$), $k\geq 0$ is an index and for each $r_i$, $r_i \in \mathbb{R}^+_0$,  $P,Q, P_i, Q_{i}$ are sequential process terms, and $S$ stands for a {\dhcsp} process term, that may be parallel.
The informal meaning  of the individual constructors is as follows:
\begin{itemize}
\item $\pskip$, $x := e$, $\pwait \ d$, $ch?x$, $ch!e$, $P;Q$, $\talloblong_{i\in I} (io_i\rightarrow Q_i)$, $B \rightarrow P$, $P \sqcap Q$ and $P^*$ are defined the same as in HCSP.


\item $\evolutiondde{F}{\s}{B}$ is the time-delay continuous evolution statement. It forces the vector $\s$ of real variables to obey the DDE $F$ as long as $B$, which defines the domain of $\s$, holds, and terminates when $B$ turns false. Without loss of generality, we assume that the set of $B$ is open, thus the escaping point will be at the boundary of $B$.
    The special case when $k=0$ corresponds to an ODE that models continuous evolution without delay.
The communication interrupt $\exempt{\evolutiondde{F}{\s}{B}}{i\in I}{io_i}{Q_i}$ behaves like $\evolutiondde{F}{\s}{B}$, except that the continuous evolution is preempted as soon as one of the communications $io_{i}$ takes place, which is followed by the respective $Q_{i}$. These two statements are the essential extensions of {\dhcsp} from HCSP.

\item For $n \ge 1$, $P_1 \| P_2 \| \ldots \| P_n$ builds a system in which $n$ concurrent processes run independently and communicate with each other along the common channels connecting them.
\end{itemize}

To better understand {\dhcsp}, we introduce delay behavior to the water tank system considered in \cite{Ahmad14,Yan16}.
\begin{example} \label{eg:wts-1}
The system is a parallel composition of two components \emph{Watertank} and \emph{Controller}, modelled by \emph{WTS} as follows:
{\small
\begin{eqnarray*}
\textit{WTS} &\Define& \textit{Watertank} \| \textit{Controller} \\
\textit{Watertank} &\Define& v:=v_0; d:=d_0; (v=1 \to \\ & &
\langle \dot{d}(t)=Q_{max} - \pi s^2  \sqrt{g (d(t)+d(t-r))}\rangle \trianglerighteq (wl!d \to cv?v); \\
& &  v=0 \to \langle \dot{d}(t)=- \pi  s^2  \sqrt{g (d(t)+d(t-r))}\rangle \trianglerighteq (wl!d \to cv?v))^* \\
\textit{Controller} &\Define& y:=v_0;x:=d_0;(\pwait \ p; wl?x; \\ & & x\ge \textit{ub} \to y:=0;x \le \textit{lb} \to y:=1; cv!y)^*
\end{eqnarray*}
}
where $Q_{max}$, $\pi$, $s$ and $g$ are system parameters, the control variable $v$ can take two values, $1$ or $0$, which indicate the watering valve on the top of the tank is open or closed, respectively, $d$ is the water level of the $\textit{Watertank}$ and its dynamics depends on the value of $v$.
 For each case, the evolution of $d$ follows a DDE that is governed by both the current state and the past state $r$ time ago. The time delay $r$  accounts for time involved in communication between the watertank and the controller.

The system is initialized by an initial state, i.e., $v_0$ and $d_0$ for the controller variable and water level, respectively. $wl$ and $cv$ are channels connecting $\textit{Watertank}$ and $\textit{Controller}$ for transferring information (water level and control variable respectively) between them. In the \emph{Controller}, the control variable $y$ is updated with a period of $p$, and its value is decided by the water level read from the $\textit{Watertank}$ ($x$ in \emph{Controller}). If $x\ge \textit{ub}$ holds, where $\textit{ub}$ is an upper bound, $y$ is set to $0$ (valve closed), else if $x \le \textit{lb}$ holds, where $\textit{lb}$ is a lower bound, $y$ is set to $1$ (valve open), otherwise, $y$ keeps unchanged. Basically, starting from the initial state, $\textit{Watertank}$ and $\textit{Controller}$ run  independently for $p$ time, then $\textit{Watertank}$ sends the current water level to $\textit{Controller}$, according to which the value of the control variable is updated and then sent back to $\textit{Watertank}$, after that, a new period repeats. The goal of the system is to maintain the water level within a desired scope.
\end{example}
\oomit{A formal operational semantics for {\dhcsp} with labelled transition system can be easily given similarly for
HCSP, please refer to \cite{Yan16,Zhan16}.}

\subsection{Semantics of {\dhcsp}}
\label{section:semanticsofdhcsp}
In order to define an operational semantics of {\dhcsp}, we use non-negative reals \sm{\RR^+}  to model time, and introduce a global clock $\textit{now}$   to record the time in the execution of a process. Different from ODE, the solution of a DDE at a given time
is not a single value, but a time function. Thus, to interpret a process $S$,   we first define a  state $\staten$ as the following mapping:
\[\begin{array}{lll}
 \staten:  (\variable(S)  \to (\Intv \to \RR^n)) \cup  (\{\textit{now}\} \rightarrow \RR^+)
\end{array}
  \]
  where $\variable(S)$ represents the set of state variables of $S$, and $\Intv$ is a timed interval.
 The semantics of each state variable with respect to a state is defined as a mapping from a timed interval to the value set.
  We denote  by $\mathcal{D}$ the set of such states.
 In addition, we introduce  a flow $H$ as a mapping from a timed interval to a state set, i.e.
 $H: \Intv \to \mathcal{D}$
 called \emph{flow}, to represent the continuous flow of process $S$ over the timed interval  $\Intv$.

  A structural operational semantics of {\dhcsp} is defined by a set of transition rules. Each transition rule has the form of $(P, \rho) \leadm{\alpha} (P', \rho', H)$, where $P$ and $P'$ are {\dhcsp} processes, $\alpha$ is an event, $\rho, \rho'$ are  states, $H$ is a  \emph{flow}. It expresses that, starting from initial state $\rho$, by
performing event $\alpha$, $P$ evolves into $P'$, ends in state $\rho'$, and produces the execution flow $H$.   The label $\alpha$ represents events, which can be a discrete non-communication event, e.g. skip, assignment, or the evaluation of Boolean expressions, uniformly denoted by $\tau$, or an external communication event $ch!c$ or $ch?c$, or an internal communication $ch. c$, or a time delay $d$, where $c \in \RR, d\in \RR^+$. When both $ch!c$ and $ch?c$ occur, a communication $ch. c$ occurs.

Before defining the semantics, we introduce an abbreviation for manipulating states.
Given a state $\rho$, $d \in \RR^+$, and a set of variables $V$, $\rho[V\Downarrow_d]$ means the clock takes progress for $d$ time units, and the values of the variables in $V$ at time $\rho(\textit{now})+d$ is defined as a constant function over timed interval $[\rho(\textit{now}), \rho(\textit{now})+d]$. Precisely, for any $t$ in the domain,
\[\rho[V\Downarrow_d](x)(t) \Define \left\{
\begin{array}{lll}
   \rho(x)(t)& \mbox{if $x \notin V$}\\
    \rho(x)(\rho(\textit{now})) & \mbox{otherwise}
\end{array}
\right. \]

For space of limitation, we only present the transition rules for the time-delayed continuous evolution statement here,
the rules for other constructors can be defined similarly to the ones in HCSP, see \cite{Zhan16}.
The first rule  represents that the DDE evolves for $d$ time units, while $B$ always preserves true throughout the
extended interval.
{\small \[
   \fracN{\begin{array}{c}
	 \mbox{Assume $X :[0, \infty)\to ([-r, \infty] \to \mathbb{R}^{d(\s)})$ is the solution of  $\evolutionddesem{F}{\s}{B}$}\\
	   \mbox{  with initial value $\s(t)= H(t)(\s)(t)$ for $t \in [\rho(\now) - r, \rho(\now)]$ and}  \\
 \forall d >0. \forall t\in[0,d), \newsemhh{B}{\rho[\now\mapsto \now+t, \s \mapsto X_t] }  = \ptrue
	   \end{array}}
	  {(\evolutionddesem{F}{\s}{B},
	    \rho) \leadm{d}
	   \left( \begin{array}{l} \evolutionddesem{F}{\s}{B}, \\
	      \rho[V\backslash\{\s\}\Downarrow_d][\now \mapsto \now+d,
	    \s \mapsto X_d],
	      H_d^{\rho, \s, X}
	     \end{array} \right)}
\] }
 where $H$ is the initial history before executing the DDE (recording the past state of $\s$); and for any $t$, $X_t$ is defined as a function over timed interval $[\rho(\now), \rho(\now) +t]$
   such that $X_t (a) = X(t)(a-\rho(\now))$ for each $a$ in the domain; and
   the produced flow $H_d^{\rho, \s, X}$ is defined as:  $\forall t \in [\rho(\now), \rho(\now)+d]. H_d^{\rho, \s, X}(t) = \rho[\now \mapsto t, \s\mapsto X_{t-\rho(\now)}]$.

The second rule represents that, when the negation $\neg B$  is true at the initial state, the DDE terminates.
\[
 \fracN{ \begin{array}{l}
	
	   \newsemhh{ \neg \textit{B} }{\rho}  = \ptrue
	  \end{array}}
	  {\begin{array}{l} (\evolutionddesem{F}{\s}{B}, \rho) \leadm{\tau}
	   (\pstop, \rho)
	     \end{array}}
\]

\subsection{Approximate Bisimulation on \emph{d}HCSP}
\label{section:appbisimofTS}

First of all, as a convention, we use
$\compress{}{\alpha}{}$ to denote the $\tau$ transition closure of transition $\alpha$, i.e., there is a sequence of $\tau$ actions before and/or after $\alpha$.
Given a state $\rho$ defined over interval $[t_1, t_2]$, for each $t \in [t_1, t_2]$, we define $\rho \downharpoonright _t$ of type $\variable(S)\cup \{\now\}  \to \textit{Val}$ to restrict the value of each variable to the
result of the corresponding function at time $t$:
\[\rho \downharpoonright _t (x) = \left\{
\begin{array}{ll}
   \rho(x)(t) &\mbox{ for all $x \in \variable(S)$}\\
   \rho(x) &\mbox{for $x=\now$}
\end{array}\right.\]
With this function, we can reduce the operations manipulating a state with function values to the ones manipulating states with point values.
Meanwhile, we assume $\compress{(S,\rho) }{0}{ (S,\rho)}$
always holds for any process $S$ and state $\rho$.

\begin{definition}[Approximate bisimulation] Suppose $\mathcal{B}$ is a symmetric binary relation on {\dhcsp} processes
such that $S_1$ and $S_2$ share the same set of state variables for $(S_1,S_2)\in \mathcal{B}$, and $\textbf{d}$
is the metric of $L^2$ norm, and  $h\in \Bbb{R}^+$ and $\varepsilon\in \Bbb{R}^+$
are the given time and value precision, respectively. Then, we say
$\mathcal{B}$ is an approximately bisimulation w.r.t. $h$ and $\varepsilon$, denoted by $\mathcal{B}_{h,\varepsilon}$, if
 for any $(S_1,S_2)\in \mathcal{B}_{h,\varepsilon} $, and
 $(\rho_1,\rho_2)$ with $\textbf{d}(\rho_1\downharpoonright _{\rho_1(\now)},\rho_2\downharpoonright _{\rho_2(\now)}) \leq \varepsilon$, the following conditions are satisfied:
\begin{itemize}
\item[1.] if $\compress{(S_1,\rho_1) }{\alpha}{ (S_1^{\prime},\rho_1')}$ and $\alpha \notin \RR^+$, then  there exists
    $(S_2',\rho_2')$ such that
     $\compress{(S_2,\rho_2) }{\alpha}{(S_2^{\prime},\rho_2')}$, $(S_1',S_2')\in \mathcal{B}_{h,\varepsilon} $ and
 $\textbf{d}(\rho_1'\downharpoonright _{\rho_1'(\now)},\rho_2'\downharpoonright _{\rho_2'(\now)}) \leq \varepsilon$,
  or
    there exist
    $(S_2^*,\rho_2^*)$, $(S_2',\rho_2')$ and $0<t\leq h$ such that
     $\compress{(S_2,\rho_2) }{t}{(S_2^{*},\rho_2^*, H_2^*)}$,
     $\compress{(S_2^*,\rho_2^*) }{\alpha}{(S_2^{\prime},\rho_2')}$,
      $(S_1,S_2^*)\in \mathcal{B}_{h,\varepsilon}$,
      $(S_1',S_2')\in \mathcal{B}_{h,\varepsilon} $ and
 $\textbf{d}(\rho_1'\downharpoonright _{\rho_1'(\now)},\rho_2'\downharpoonright _{\rho_2'(\now)}) \leq \varepsilon$.

 \item[2.] if $\compress{(S_1,\rho_1) }{t}{ (S_1^{\prime},\rho_1', H_1)}$ for some $t>0$, then there exist
    $(S_2',\rho_2')$ and $t'\geq 0$ such that
     $|t-t'| \leq h$,
     $\compress{(S_2,\rho_2) }{t'}{(S_2^{\prime},\rho_2', H_2)}$, $(S_1',S_2')\in \mathcal{B}_{h,\varepsilon}$, and for any $o \in [\rho(\textit{now}), \rho(\textit{now})+\min (t,t')]$,
  $\textbf{d}(\rho_1'\downharpoonright _o,\rho_2'\downharpoonright _o) \leq \varepsilon$;
  and  for any $o \in [\rho(\textit{now})+\min (t,t'), \rho(\textit{now})+\max (t,t')]$, $\textbf{d}(\rho_1'\downharpoonright _{o_1},\rho_2'\downharpoonright _{o_2}) \leq \varepsilon$ where $o_1 = \min(o, \rho(\textit{now})+ t)$ and $o_2 = \min(o, \rho(\textit{now})+ t')$.

\end{itemize}
\label{definition:appbisimulation}
\end{definition}
\oomit{
\begin{definition}[Approximate bisimulation] Suppose $\mathcal{B}$ is a symmetric binary relation on {\dhcsp} processes
such that $S_1$ and $S_2$ share the same set of state variables for $(S_1,S_2)\in R$, and $\textbf{d}$
is the metric of $L^2$ norm, and  $h\in \Bbb{R}^+$ and $\varepsilon\in \Bbb{R}^+$
are the given time and value precision, respectively. Then, we say
$\mathcal{B}$ is an approximately bisimulation w.r.t. $h$ and $\varepsilon$, denoted by $\mathcal{B}_{h,\varepsilon}$, if
 for any $(S_1,S_2)\in \mathcal{B}_{h,\varepsilon} $, and
 $(\sigma_1,\sigma_2)$ with $\textbf{d}(\sigma_1,\sigma_2) \leq \varepsilon$, the following conditions are satisfied:
\begin{itemize}
\item if $\compress{(S_1,\sigma_1) }{l}{ (S_1^{\prime},\sigma_1')}$, then either there exists
    $(S_2',\sigma_2')$ such that
     $\compress{(S_2,\sigma_2) }{l}{(S_2^{\prime},\sigma_2')}$, $(S_1',S_2')\in \mathcal{B}_{h,\varepsilon} $ and
 $\textbf{d}(\sigma_1',\sigma_2') \leq \varepsilon$, or
    there exist
    $(S_2^*,\sigma_2^*)$, $(S_2',\sigma_2')$ and $0<t\leq h$ such that
     $\compress{(S_2,\sigma_2) }{t}{(S_2^{*},\sigma_2^*, H_2^*)}$,
     $\compress{(S_2^*,\sigma_2^*) }{l}{(S_2^{\prime},\sigma_2')}$,
      $(S_1,S_2^*)\in \mathcal{B}_{h,\varepsilon}$,
      $(S_1',S_2')\in \mathcal{B}_{h,\varepsilon} $ and
 $\textbf{d}(\sigma_1',\sigma_2') \leq \varepsilon$.

 \item if $\compress{(S_1,\sigma_1) }{t}{ (S_1^{\prime},\sigma_1', H_1)}$ for some $t>0$, then there exist
    $(S_2',\sigma_2')$ and $t'>0$ such that
     $|t-t'| \leq h$,
     $\compress{(S_2,\sigma_2) }{t'}{(S_2^{\prime},\sigma_2', H_2)}$, $(S_1',S_2')\in \mathcal{B}_{h,\varepsilon}$,
 $\textbf{d}(H_1(\tau),H_2(\tau)) \leq \varepsilon$ for any $\tau \in [\textit{now}, \textit{now}+\min (t,t')]$ and
 $\textbf{d}(\sigma_1',\sigma_2') \leq \varepsilon$.
\end{itemize}
\label{definition:appbisimulation}
\end{definition}
}

\begin{definition} Two {\dhcsp} process $S_1$ and $S_2$
 are approximately bisimilar with respect to precision $h$ and $\varepsilon$, denoted by $S_{1}\cong_{h, \varepsilon} S_{2}$, if there exists  an $(h, \varepsilon)$-approximate bisimulation relation $\mathcal{B}_{h,\varepsilon}$  s.t.
  $(S_1, S_2) \in \mathcal{B}_{h,\varepsilon}$.
\label{definition:TSappbisimulation}
\end{definition}

\begin{theorem}
  \label{theorem:decidable}
  Given two {\dhcsp} processes, it is decidable whether they are approximately bisimilar on $[0,T]$ for a given $T\in \mathbb{R}^+$.
\end{theorem}

\oomit{
\subsection{Approximate Bisimulation on {\dhcsp}}
\label{section:appbisimofhcsppro}
In this paper, we will define the semantics of the source {\dhcsp} model and the target discretized model based on the notion of transition system, and then observe the approximate bisimulation relation between them. As the process on how to derive the transition system from a {\dhcsp} process is nearly the same as that from an HCSP process \cite{Yan16}, we will not explain it here.
The definition of approximate bisimulation between two {\dhcsp} processes is given below.
\begin{definition}
Let $S_1$ and $S_2$ be two {\dhcsp} processes, $T\in \mathbb{R}^+$ be the upper bound of the execution time, and $h, \varepsilon$ the time and value precisions. Let $v_0$ be an arbitrary initial state. $S_1$ and $S_2$ are $(h, \varepsilon)$-\emph{approximately bisimilar} on $[0,T]$, denoted by $S_1\cong_{h, \varepsilon}S_2$ on $[0,T]$, if $TS(S_1)\cong_{h, \varepsilon}TS(S_2)$, in which $TS(S_1)$ and $TS(S_2)$ are the $\tau$-compressed transition systems of $S_1$ and $S_2$ with the same initial state $v_0$ within execution time $T$, respectively.
\end{definition}

For a given $T\in \mathbb{R}^+$, the approximate bisimulation relation between $S_1$ and $S_2$ needs to be satisfied on $[0,T]$. As the execution time $T$ of processes  $S_1$ and $S_2$ is bounded, plus that $S_1$ and $S_2$ can only perform finite number of discrete actions, the transition systems derived from $S_1$ and $S_2$ (i.e., $TS(S_1)$ and $TS(S_2)$ respectively) are \emph{symbolic}. By adopting Alg. 3 in \cite{Girard07a}, we can obtain the maximal (coarsest) approximate bisimulation relation between $TS(S_1)$ and $TS(S_2)$, then whether $TS(S_1)$ and $TS(S_2)$ are approximately bisimilar can be decided by inspecting whether their initial states are all in the maximal approximate bisimulation relation (from Def. \ref{definition:TSappbisimulation}). Therefore, we can conclude that whether $S_1\cong_{h, \varepsilon}S_2$ on $[0,T]$ is decidable.
}

\section{Discretization of {\dhcsp}}
\label{section:discretizationofhcsp}

The process on generating code from {\dhcsp} is similar to that from HCSP~\cite{Yan16}, consisting of
two phases: (1) discretization of the {\dhcsp} model; (2) code generation from the discretized {\dhcsp} model to SystemC.

Benefiting from its compositionality, {\dhcsp} can be discretized by defining rules for all the constructors, in  which the discretization of  delay continuous dynamics (i.e., DDE) is most critical. Let $S$ be a {\dhcsp} process, $T\in \mathbb{R}^+$ be a time bound, $ h $ and $\varepsilon$ be the given precisions for time and value, respectively. Our goal is to construct a discrete {\dhcsp}
process $D_{h,\vare}(S)$ from $S$, s.t.  $S$ is $(h,\varepsilon)$-approximately bisimilar to $D_{h,\vare}(S)$ on $[0,T]$, i.e., $\bisimilar{S}{D_{h,\vare}(S)}$ on $[0,T]$.
To achieve this, 
we firstly introduce a simulation-based method (inspired by \cite{Chen16}) for discretizing a single DDE and then extend it for multiple DDEs to be executed in sequence; afterwards, we present the discretization of {\dhcsp} in bounded time.

\subsection{Discretization of DDE (DDEs) in Bounded Time}
\label{section:Dis ConD}
 To solve DDEs  is much more difficult than to solve ODEs, as DDEs are history dependent, therefore, non-Markovian, in contrast,
  ODEs are history independent and Markovian. So, in most cases, explicit solutions to DDEs are impossible, therefore, DDEs are normally solved by using approximation based techniques \cite{Bellen2013}.  In \cite{Chen16}, the authors propose a novel method for safety verification of delayed differential dynamics, in which a validated simulator for a DDE is presented. The simulator produces a sequence of discrete states for approximating the trajectory of a DDE and meanwhile calculates the corresponding local error bounds. Based on this work, we can obtain a validated discretization of a DDE w.r.t. the given precisions $h$ and $\varepsilon$. Furthermore, we can easily extend the simulator to deal with systems containing multiple DDEs in sequence.

Next we first consider the discretization of a DDE within bounded time $T_d\in \RR^+$, for some $T_d\leq T$. The purpose is to find a discrete step size $h$ s.t. the DDE and its discretization are $(h,\euler)$-approximately bisimilar within $[0,T_d]$, for a given precision $\euler$ that is less than the global error $\varepsilon$.
For simplifying the notations, we consider a special case of DDE in which only one delay term, $r>0$, exists, as in
\begin{equation}
\label{eq:dynamical-special}
\left\{
\begin{array}{ll}
\dot{\xx}(t)=\ff(\xx(t),\xx(t-r)),
& t\in [0,\infty ) \\
\xx(t)=\fg(t), & t\in [-r,0]
\end{array}\right.
\end{equation}
where we use $\ff(\xx,\xx_r)$ to denote the dynamics, $\xx$ for the current state and $\xx_r$ for the past state at $t-r$. In fact, the method for this special case can be easily extended to the general case as in (\ref{eq:dynamical}), by recording the past states between $t-r_k$ and $t$, the detailed discussion can be found in \cite{Chen16}.

For a DDE $\ff(\xx,\xx_r)$ with initial condition $\fg(t)$ which is continuous on $[-r,0]$, delay term $r$, step size $h$, and time bound $T_d$, the validated simulator in \cite{Chen16} can produce three \emph{lists} (denoted as $\Lbrack \cdot \Rbrack $) with the same length, namely, (1) $\ttt=\Lbrack t_{-m},...,t_0,...,t_n \Rbrack $, storing a sequence of time stamps on which the approximations are computed ($t_{-m},...,t_{0}$ for the time before $0$, i.e., $[-r,0]$, with $m=r/h$), satisfying  $t_{-m},...,t_{-1}<0=t_0<t_1<...<t_n=T_d$ and $t_i-t_{i-1}=h$ for all $i\in[-m+1,n]$, (2) $\yy=\Lbrack \xx_{-m},...,\xx_0,\xx_1,...,\xx_n \Rbrack $, recording a sequence of approximate states of $\xx$ starting from $\xx_{-m}$, corresponding to time stamps in $\ttt$, (3) $\dd=\Lbrack d_{-m},...,d_0,d_1,...,d_n \Rbrack $, recording the corresponding sequence of local error bounds. The implementation of the simulator is based on the well-known \emph{forward Euler method}, i.e., $\xx:= \xx+h\ff(\xx,\xx_r)$. In addition, we usually require the delay term $r$ be an integral multiple of the step size $h$, i.e., $m \in \mathbb{N}^+$, in order to ensure the past state $\xx_r$ could be found in $\yy$.

A remarkable property of the simulator
\[
X(t,\fg(0))\in \textit{conv}(N(\xx_i,d_i)\cup N(\xx_{i+1},d_{i+1}))
\]
holds for each $t\in[t_i,t_{i+1}]$ with $i=0,1,...,n-1$, where $X(\cdot)$ is the trajectory  of $\dot{\xx}=\ff(\xx,\xx_r)$, and $N(\xx_i,d_i)$ is the $d_i$-neighbourhood of $\xx_i$ ($\xx_i$ and $d_i$ are elements of $\yy$ and $\dd$, respectively). Based on this fact, we can use $\xx_{i+1}$ as the approximation of $X(t,\fg(0))$ for all $t\in[t_i,t_{i+1}]$ for any $i\in[0,n-1]$, s.t. the DDE (\ref{eq:dynamical-special}) and the sequence $\Lbrack \xx_0,\xx_1,...,\xx_n \Rbrack $ are $(h,\euler)$-approximately bisimilar on $[0,T_d]$, if the diameter of every $\textit{conv}(N(\xx_i,d_i)\cup N(\xx_{i+1},d_{i+1}))$ is not greater than the precision $\euler$, i.e., $\textit{dia}(\textit{conv}(N(\xx_i,d_i)\cup N(\xx_{i+1},d_{i+1})))<\euler$ for all $i\in[0,n-1]$.

\begin{theorem}[Approximation of a DDE] Let $\Gamma$ be a DDE as in (\ref{eq:dynamical-special}), and $\ff$ in (\ref{eq:dynamical-special}) is continuously differentiable on $[0,T_d]$, and $\xx_0 \in \RR^n$ with $\|\xx_0 - \fg(0)\| \le d_0$. Then for any precision $\euler>0$ and $0<d_0 <\euler$, there exists a step size $h>0$ s.t. $\Gamma$ and
\[
 \begin{array}{l}
 \xx:=\xx_0; (\pwait\ h; \xx:= \xx+h\ff(\xx,\xx_r))^{\frac{T_d}{h}};
  \end{array}
 \]
are $(h,\euler)$-approximately bisimilar on $[0,T_d]$.
\label{theorem:DDEDiscrete}
\end{theorem}

 Based on the simulation algorithm given in \cite{Chen16}, we design a method for automatically computing a step size $h$ s.t. the DDE as in (\ref{eq:dynamical-special}) and its discretization are $(h,\euler)$-approximately bisimilar on $[0,T_d]$,
    as presented in Alg. \ref{alg:compt-h-one-dde} and Alg. \ref{alg:check-h-one-dde}.

{\small
\begin{algorithm}[htb]
\caption{ComStepsize\_oneDDE: computing the step size $h$ for the one DDE}
\begin{algorithmic}[1]
\REQUIRE ~~~~
   The dynamics $\ff(\xx,\xx_r)$, initial state $\xx_0$, delay term $r$, precision $\euler$, and time bound $T_d$;
\STATE $h=r$; $v=true$; $\ttt = \Lbrack -h,0 \Rbrack $; $\yy = \Lbrack \xx_0,\xx_0 \Rbrack $; $\dd = \Lbrack 0,0 \Rbrack $;
\WHILE {\textit{true}}
\STATE $\textit{CheckStepsize}(\ff(\xx,\xx_r),r,h,\euler,[0,T_d],\ttt,\yy,\dd,v)$;
\IF {$v=\textit{false}$}
\STATE $h=h/2$; $v=\textit{true}$;
\STATE $\ttt = \Lbrack -h,0 \Rbrack $;
\ELSE
\STATE break;
\ENDIF
\ENDWHILE
\STATE return $h$;
\end{algorithmic}\label{alg:compt-h-one-dde}
\end{algorithm}
}

Alg. \ref{alg:compt-h-one-dde} is designed for computing a valid step size $h$ for a given DDE. It first initializes the value of $h$  to $r$ and Boolean variable $v$, which indicates whether the current $h$ is a valid step size, to $\textit{true}$, and the lists for simulating the DDE, i.e., $\ttt$, $\yy$, and $\dd$ (line 1). Here, we assume the initial condition is a constant function, i.e., $\xx_{t}=\xx_0$, on $[-r,0]$, therefore, states before time $0$ is represented as one state at $-h$. Then, it iteratively checks whether the current value of $h$ can make Theorem \ref{theorem:DDEDiscrete} hold, by calling the function \textit{CheckStepsize} that is defined in Alg. \ref{alg:check-h-one-dde} (lines 2-10). If current $h$ is not valid ($v$ is set to \textit{false} for this case), $h$ is set to a smaller value, i.e., $h/2$, and $v$ is reset to \textit{true}, and $\ttt$ is  reinitialized according to the new $h$ (lines 4-6). Otherwise,   a valid $h$ is found, then the while loop exits (lines 7-9). The termination of the algorithm can be guaranteed by Theorem \ref{theorem:DDEDiscrete}, thus  a valid $h$ can always be found and returned (line 11).
{\small
\begin{algorithm}[htb]
\caption{CheckStepsize: checking whether the step size $h$ is valid for  precision $\euler$}
\begin{algorithmic}[1]
\REQUIRE ~~~~
   The dynamics $\ff(\xx,\xx_r)$, delay term $r$, step size $h$, precision $\euler$, time span $[T_1,T_2]$, boolean variable $v$, and simulation history $\langle \ttt,\yy,\dd \rangle$ before $T_1$;
\STATE $n=length(\ttt)$; $m = r/h$;
\WHILE {$\ttt(n)<T_2$}
\STATE $\ttt(n+1)= \ttt(n)+h$;
\STATE $\yy(n+1)= \yy(n)+\ff(\yy(n),\yy(n-m))*h$;
\STATE $\ee(n)=$ \textbf{Find} minimum $e$ s.t.\\
$$\left\{
\begin{aligned}
&\| \ff(\xx+t*\ff,\xx_r+t*\mathbf{g})-\ff(\yy(n),\yy(n-m)) \| \le e-\sigma, for  \\
&\forall t\in[0,h]  \\
&\forall \xx\in N(\yy(n), \dd(n)) \\
&\forall \xx_r\in N(\yy(n-m), \dd(n-m)) \\
&\forall \ff\in N(\ff(\yy(n),\yy(n-m)), e) \\
&\forall \mathbf{g}\in N(\ff(\yy(n-m),\yy(n-2m)), \ee(n-m));
\end{aligned}
\right.
$$
\STATE $\dd(n+1)= \dd(n)+h*\ee(n)$;
\IF {$\max(\yy(n)+\dd(n),\yy(n+1)+\dd(n+1))- \min(\yy(n)-\dd(n),\yy(n+1)-\dd(n+1))>\euler$}
\STATE $v=\textit{false}$;
\STATE break;
\ELSE
\STATE $\ttt = \Lbrack \ttt,\ttt(n+1) \Rbrack $; $\yy = \Lbrack \yy,\yy(n+1) \Rbrack $; $\dd = \Lbrack \dd,\dd(n+1) \Rbrack $;;
\STATE $n = n+1$;
\ENDIF
\ENDWHILE
\STATE return $\langle v, \ttt,\yy,\dd \rangle$;
\end{algorithmic}\label{alg:check-h-one-dde}
\end{algorithm}
}

Alg. \ref{alg:check-h-one-dde} implements function \textit{CheckStepsize},  which is slightly different from the simulation algorithm given in \cite{Chen16}. The history of $\langle \ttt,\yy,\dd \rangle$ is added to the inputs,  for simulating multiple DDEs in sequence.
At the beginning, the variable $n$ that stores the last recent simulation step is initialized as the length of  current $\ttt$, and an offset $m$ is set to $r/h$ thus $\yy(n-m)$, i.e., the $(n-m)$th element of list $\yy$, locates the delayed approximation at time $\ttt(n)-r$ (line 1). When current time (i.e., $\ttt(n)$) is less than the end of the time span (i.e., $T_2$),  the lists $\ttt$, $\yy$ and $\dd$ are iteratively updated by adding new elements, until $T_2$ is reached (lines 2-14).
In each iteration, firstly, the time stamp is added by the step size $h$ and the approximate state at this time is computed by the \emph{forward Euler method} (line 4), and then the local error bound $\dd(n+1)$ is derived based on the local error slope $\ee(n)$ (line 6), which is reduced to a constrained optimization problem (line 5) that can be solved by some solvers in Matlab or
by some SMT solvers like iSAT \cite{FraenzleEA:iSAT:JSAT} which can return a validated result, please refer to \cite{Chen16} for the details. After these values are computed, whether the diameter of the convex hull of the two adjacent approximate points at the time stamps $\ttt(n)$ and $\ttt(n+1)$  by taking their local error bounds into account greater than the given error $\euler$ is checked (lines 7-13).
If the diameter is greater than $\euler$, the while loop is broken and $v$ is set to \textit{false} (lines 8-9), which means $h$ will be reset to $h/2$ in Alg. \ref{alg:compt-h-one-dde}. Otherwise, $h$ is valid for this simulation step and the new values of $\ttt$, $\yy$ and $\dd$ are added into the  corresponding lists (lines 10-12), then a new iteration restarts  until $T_2$ is reached.  At last, the new values of $v$, $\ttt$, $\yy$ and $\dd$ are returned (line 15).

A {\dhcsp} may contain multiple DDEs, especially for those to be executed in sequence in which
 the initial states of following DDEs may depend on the flows of previous DDEs. In order to handle such cases, we present Alg. \ref{alg:compt-h-multi-dde} for computing the global step size that meets the required precision $\euler$ within bounded time $T_d$.
Suppose a sequence of DDEs $\ff_1(\xx,\xx_r),\ff_2(\xx,\xx_r),\cdots,
 \ff_k(\xx,\xx_r)$ is to be executed in sequence.
  For simplicity, assume all DDEs share the same delay term $r$, and the execution sequence of the DDEs is decided by a scheduler (\textit{Schedule} in line 6).
At the beginning, $h$ and $v$ are initialized as the delay term $r$ and \textit{true} respectively (line 1).
Then, before the current time (i.e., $\ttt(end)$) reaches the end of the time span (i.e., $T_d$), a while loop
is executed to check
whether $h$  satisfies the precision $\euler$, in which \textit{ComStepsize\_oneDDE} and \textit{CheckStepsize} are called (lines 2-13). In each iteration, the three lists $\ttt$, $\yy$ and $\dd$ are initialised as before (line 3),  then  the valid $h$ for the first DDE $\ff_1(\xx,\xx_r)$ is computed by calling \textit{ComStepsize\_oneDDE} (line 4), where $t_1$ denotes the length of the execution time of $\ff_1(\xx,\xx_r)$.
Afterwards, for the following DDEs, an inner while loop to check whether the calculated $h$ is within the error bound
 $\euler$ is executed ~¡¡(lines 5-12). Thereof, which DDE should be executed is determined
  by \textit{Schedule} (one DDE may be executed for multiple times), and the corresponding span of execution time is represented as $[t_{i-1},t_i]$ for the $i$-th DDE  (lines 6-7).    If  $h$ is not valid for some DDE, i.e., $v=\textit{false}$ (line 8),
   depending on the return value of \textit{CheckStepsize} function, a new smaller $h$ (i.e., $h/2$) is chosen and
   $v$ is reset to \textit{true}, then the inner \textit{while} loop is broken (lines 8-11) and a new iteration restarts from time $0$ with the new $h$ (line 3); Otherwise, a valid $h$ is found (line 13).
      Since we can always find small enough step size to make all DDEs meet the precision within $[0,T_d]$ by Theorem \ref{theorem:DDEDiscrete}, Alg. \ref{alg:compt-h-multi-dde} is ensured to terminate (line 14).

{\small
\begin{algorithm}[htb]
\caption{ComStepsize\_multiDDEs: computing the step size $h$ for multiple DDEs}
\begin{algorithmic}[1]
\REQUIRE ~~~~
   A sequence of dynamics $\ff_1(\xx,\xx_r),\ff_2(\xx,\xx_r),...,\ff_k(\xx,\xx_r)$, initial state $\xx_0$, delay term $r$, precision $\euler$, and time bound $T_d$ (assume running from $\ff_1(\xx,\xx_r)$);
\STATE $h=r$; $v =\textit{true}$;
\WHILE {$\ttt(\textit{end})<T_d$}
\STATE $\ttt = \Lbrack -h,0 \Rbrack $; $\yy = \Lbrack \xx_0,\xx_0 \Rbrack $; $\dd = \Lbrack 0,0 \Rbrack $;
\STATE $h=\textit{ComStepsize\_oneDDE}(\ff_1(\xx,\xx_r),\xx_0,r,\euler,t_1)$;

\WHILE {$\ttt(end)<T_d$}
\STATE $i=\textit{Schedule}(\ff_1(\xx,\xx_r),\ff_2(\xx,\xx_r),...,\ff_k(\xx,\xx_r))$;
\STATE $\textit{CheckStepsize}(\ff_i(\xx,\xx_r),r,h,\euler,[t_{i-1},t_i],\ttt,\yy,\dd,v)$;
\IF {$v=\textit{false}$}
\STATE $h=h/2$; $v =\textit{true}$;
\STATE break;
\ENDIF
\ENDWHILE

\ENDWHILE
\STATE return $h$;
\end{algorithmic}\label{alg:compt-h-multi-dde}
\end{algorithm}
}
\subsection{Discretization of {\dhcsp} in Bounded Time}
\label{section:Discretizationhcsp}
Based on the above fact, we can define a set of rules to discretize a given {\dhcsp} process $S$ and obtain a discrete {\dhcsp} process
$\discrete{h}{\varepsilon}{S}$ such that they are $(h,\varepsilon)$-approximately bisimilar on $[0,T]$, for given $h$, $\varepsilon$ and $T$.
The rule for the discretization of DDE is given below, and other rules are same as the ones for
 HCSP presented in \cite{Yan16}.

\[ \fracN{\evolutionn{\dot{\xx}=\ff(\xx,\xx_r)}{B}}{
\begin{array}{c}
(N(B,\varepsilon)\wedge N^{\prime}(B,\varepsilon) \rightarrow (\pwait \ h; \xx:=\xx+h \ff(\xx,\xx_r)))^{\frac{T}{h}}; \\
N(B, \varepsilon)\wedge N^{\prime}(B,\varepsilon) \rightarrow \nstop
\end{array}
}\]

For a Boolean expression $B$, $N(B, \varepsilon)$ is defined as its $\varepsilon$-neighbourhood. For instance, $N(B, \varepsilon)=\{x|x>2-\varepsilon\}$ for $B=\{x|x>2\}$.
Then,
$\evolutionn{\dot{\xx}=\ff(\xx,\xx_r)}{B}$ is discretized as follows: 
first, execute a sequence of assignments ($T/h$ times) to $\xx$ according to \textit{Euler method}, i.e., $\xx:=\xx+h \ff(\xx,\xx_r)$, whenever $N(B, \varepsilon)\wedge N^{\prime}(B,\varepsilon)$ holds,
 where $N^{\prime}(B,\varepsilon)= N(B,\varepsilon)[\xx\mapsto \xx+h \ff(\xx,\xx_r)]$, i.e., the value of $N(B, \varepsilon)$ at the next discretized step; then, if both $N(B, \varepsilon)$ and $N^{\prime}(B,\varepsilon)$ still hold, but the time has already reached the upper bound $T$, the process behaves like $\nstop$, 
 which indicates that the behavior after $T$ will not be concerned.


\subsection{Correctness of the Discretization}
\label{section:app-hcsp}

In order to ensure $\discrete{h}{\varepsilon}{S}$ defined in Sec. \ref{section:Discretizationhcsp} is approximately bisimilar to $S$, we need to put some extra conditions on $S$, i.e., requiring it to be robustly safe. The  condition is similar
to that in ~\cite{Yan16}.
We define the $(-\epsilon)$-neighbourhood like the $\epsilon$-neighbourhood, i.e., for a set $\phi\subseteq  \mathbb{R}^n$ and $\epsilon\ge 0$, $N(\phi, -\epsilon)=\{ \xx | \xx \in \phi \wedge \forall \yy \in \neg \phi. \|\xx-\yy\| > \epsilon\}$. Intuitively, $\xx \in N(\phi, -\epsilon)$ means $\xx$ is inside $\phi$ and moreover the distance between it and the boundary of $\phi$ is greater than $\epsilon$. To distinguish the states of process $S$ from those of dynamical systems, we use $\rho$ ($\rho_0$ for initial state) to denote the states of $S$ here. Below, the notion of a robustly safe system is given.
\begin{definition}[$(\delta, \epsilon)$-robustly safe]
   Let $\delta>0$ and  $\epsilon>0$ be the given time and value precisions respectively. A {\dhcsp} process $S$ is $(\delta, \epsilon)$-robustly safe with respect to a given initial state $\rho_0$, if the following two conditions hold:
   \begin{itemize}
      \item for every continuous evolution $\evolutionn{\dot{\xx}=\ff(\xx,\xx_r)}{B}$ occurring in $S$, when  $S$ executes up to $\evolutionn{\dot{\xx}=\ff(\xx,\xx_r)}{B}$ at time $t$ with state $\rho$,  if $\rho(B) = \textit{false}$, and
          there exists $\widehat{t} > t$ with $\widehat{t} -t < \delta$  and
            $\dd(\rho, \rho_0[\xx \mapsto X(\widehat{t}, \rho_0(\xx)])) < \epsilon$, then  $\rho \in N(\neg B, -\epsilon)$;
    \item for every alternative process $B \rightarrow P$ occurring in $S$, if $B$ depends on continuous variables of $S$, then when $S$ executes up to $B \rightarrow P$ at state $\rho$,  $\rho \in N(B, -\epsilon)$ or $\rho\in N(\neg B, -\epsilon)$.
   \end{itemize}
\label{Def:robustly safe}
\end{definition}
Intuitively, the $(\delta, \epsilon)$-robustly safe condition ensures the difference, between the violation time of the same
Boolean condition $B$ in $S$ and $\discrete{h}{\varepsilon}{S}$, is bounded. As a result, we can choose appropriate values for $\delta, \epsilon, h$ and $\varepsilon$ s.t. $S$ and $\discrete{h}{\varepsilon}{S}$ can be guaranteed to have the same control flows, and furthermore the distance between their ``jump'' time (the moment when Boolean condition associated with them become false) can be bounded by $h$. Finally the ``approximation'' between the behavior of $S$ and $\discrete{h}{\varepsilon}{S}$ can be guaranteed.
The range of both $\delta$ and $\epsilon$ can be estimated by simulation.

Based on the above facts, we have the main theorem as below.
\begin{theorem}[Correctness]
 Let $S$ be a {\dhcsp} process and $\rho_0$ the initial state at time $0$. Assume $S$ is $(\delta, \epsilon)$-robustly safe with respect to $\rho_0$. Let $0< \varepsilon < \epsilon$ be a precision and $T\in \RR^+$ a time bound. If for any  DDE $\dot{\xx}=\ff(\xx,\xx_r)$ occurring in $S$, $\ff$ is continuously differentiable on $[0,T]$, and there exists $h$ satisfying $h<\delta<2h$ if $\delta>0$ s.t. Theorem ~\ref{theorem:DDEDiscrete} holds for all $\ff$ in $S$,
then  $S\cong_{h,\varepsilon}\discrete{h}{\varepsilon}{S}$ on $[0,T]$.
 \label{theorem:app-process}
\end{theorem}

Notice that for a given precision $\varepsilon$, there may not exist an $h$ satisfying the conditions in Theorem~\ref{theorem:app-process}.
It happens when the DDE fails to leave far enough away from the boundary of its domain $B$ in a limited time. However, for the special case that $\delta=0$, we can always find a sufficiently small $h$ such that $S\cong_{h,\varepsilon}\discrete{h}{\varepsilon}{S}$ on $[0,T]$.


\section{From Discretized {\dhcsp} to SystemC}
\label{section:codegenerationofhcsp}


For a {\dhcsp} process $S$, its discretization $\discrete{h}{\varepsilon}{S}$ is a model without continuous dynamics and therefore can be implemented with an algorithm model. In this section, we illustrate the procedure for automatically generating a piece of SystemC code, denoted as $SC(\discrete{h}{\varepsilon}{S})$, from a discretized {\dhcsp} process $\discrete{h}{\varepsilon}{S}$, and moreover ensures that they are ``equivalent'', i.e., bisimilar. As a result, for a given precision $\varepsilon$ and time bound $T$, if there exists $h$ such that Theorem~\ref{theorem:app-process} holds, i.e., $S\cong_{h,\varepsilon}\discrete{h}{\varepsilon}{S}$ on $[0,T]$, we can conclude that the generated SystemC code $SC(\discrete{h}{\varepsilon}{S})$ and the original {\dhcsp} process $S$ are $(h,\varepsilon)$-approximately bisimilar on $[0,T]$. 

%

Based on its semantics,
 a {\dhcsp} model that contains multiple parallel processes is mapped into an \emph{SC\_MODULE} in SystemC, and each parallel component
  is implemented as a thread, e.g., $\discrete{h}{\varepsilon}{P_1} \| \discrete{h}{\varepsilon}{P_2}$ is mapped into two concurrent threads, \emph{SC\_THREAD}$(SC(\discrete{h}{\varepsilon}{P_1}))$ and \emph{SC\_THREAD}$(SC(\discrete{h}{\varepsilon}{P_2}))$, respectively. For each sequential process, i.e., $\discrete{h}{\varepsilon}{P_i}$,  we define corresponding rule for transforming it into a piece of SystemC code, according to the type of $\discrete{h}{\varepsilon}{P_i}$.

In Table~\ref{table:codegenerationofHCSP}, parts of generation rules are shown for different types of the sequential process $\discrete{h}{\varepsilon}{P_i}$. For $x := e$, it is mapped into an equivalent assignment statement (i.e, $x = e$), followed by a statement $wait(\emph{SC\_ZERO\_TIME})$ for making the update valid. For $\pwait \ d$, it is straightforward mapped into a statement $\pwait(d, \emph{SC\_TU})$, where \emph{SC\_TU} is the time unit of $d$, such as \emph{SC\_SEC} (second), \emph{SC\_MS} (millisecond), \emph{SC\_US} (microsecond), etc. The sequential composition and alternative statements are defined inductively. Nondeterminism is implemented as an \emph{if-else} statement, in which $rand()\%2$ returns $0$ or $1$ randomly. A $while$ statement is used for implementing the repetition constructor, where $\repnum(P^*)$ returns the upper bound of the repeat times for
$P$.
\begin{table}[t]
\centering
\begin{tabular}{lll}
\hline
$x:=e $& $\rightarrow$ & $x = e; wait(\emph{SC\_ZERO\_TIME});$\\[0.2em]

$\pwait \ d$ & $\rightarrow$ & $wait(d, \emph{SC\_TU});$\\[0.2em]

$\discrete{h}{\varepsilon}{P}; \discrete{h}{\varepsilon}{Q}$& $\rightarrow$ & $SC(\discrete{h}{\varepsilon}{P}); SC(\discrete{h}{\varepsilon}{Q});$\\[0.2em]

$B \rightarrow \discrete{h}{\varepsilon}{P}$& $\rightarrow$ & $if (B) \{SC(\discrete{h}{\varepsilon}{P});\}$\\[0.2em]

\multirow{2}{*}{$\discrete{h}{\varepsilon}{P} \sqcap \discrete{h}{\varepsilon}{Q}$}& \multirow{2}{*}{$\rightarrow$} & $if (rand()\%2) \{SC(\discrete{h}{\varepsilon}{P});\}$\\ && $else \{SC(\discrete{h}{\varepsilon}{Q});\}$\\[0.2em]

\multirow{2}{*}{$(\discrete{h}{\varepsilon}{P})^{\ast}$}& \multirow{2}{*}{$\rightarrow$} & $while (i<=\repnum(P^*)) \{$\\ &&
$SC(\discrete{h}{\varepsilon}{P});i$++$;\}$\\[0.2em]

\hline \\
\end{tabular}
\caption{Part of rules for code generation of {\dhcsp}}
\label{table:codegenerationofHCSP}
\end{table}


In order to represent the communication statement, additional channels in SystemC (i.e., \emph{sc\_signal}) and events (i.e., \emph{sc\_event}) are introduced to ensure the synchronization between the input side and output side. Consider the discretized input statement, i.e., $ch?:=1;ch?x;ch?:=0$, Boolean variable $ch?$ is represented as an \emph{sc\_signal} (i.e., $ch\_r$) with Boolean type, and moreover additional \emph{sc\_event} (i.e., $ch\_r\_done$) is imported to represent the completion of the action that reads values from channel $ch$. As a result, the SystemC code generated from it is defined as: first, Boolean signal $ch\_r$ is initialized as $1$, which means channel $ch$ is ready for reading (lines 2-3); then, the reading process waits for the writing of the same channel from another process until it has done (lines 4-6); after that, it gets the latest value from the channel and assigns it to variable $x$ (lines 7-8); at last, it informs the termination of its reading to other processes and resets  $ch\_r$ to $0$ (lines 9-11). Here, there are two sub-phases within the second phase (lines 4-6): first, deciding whether the corresponding writing side is ready (line 4), if not (i.e., $ch\_w = 0$), the reading side keep waiting until the writing side gets ready, i.e., $ch\_w=1$ (line 5); afterwards, the reading side will wait for another event which indicates that the writing side has written a new value into the channel $ch$ (line 6), for ensuring the synchronization.
{\small
\begin{lstlisting}
// code for input statement
ch_r=1;
wait(SC_ZERO_TIME);
if(!ch_w)
    wait(ch_w.posedge_event());
wait(ch_w_done);
x=ch.read();
wait(SC_ZERO_TIME);
ch_r_done.notify();
ch_r=0;
wait(SC_ZERO_TIME);
\end{lstlisting}
}

The discretized continuous statement is mapped into two sequential parts in SystemC. For the first part, i.e., $(N(B,\varepsilon)\wedge N^{\prime}(B,\varepsilon) \rightarrow (\pwait \ h; \xx:=\xx+h \ff(\xx,\xx_r)))^{\frac{T}{h}}$, a \textbf{for} loop block is generated (lines 2-8), in which a sequence of \emph{if} statements, corresponding to Boolean condition  $(N(B,\varepsilon)\wedge N^{\prime}(B,\varepsilon)$, are executed (lines 3-7). Within every conditional statement, a \textbf{wait}
 statement and an assignment statement (based on \emph{Euler method}) are sequentially performed (lines 4-6). Here, $N(B,e)$, $N\_p(B,e)$ and $f(x,x\_r)$ are helper functions (implemented by individual functions) that are generated from $N(B,\varepsilon)$,  $N^{\prime}(B,\varepsilon)$ ($e=\varepsilon$ here)  and $\ff(\xx,\xx_r)$, respectively.
For the second part, i.e., $N(B,\varepsilon)\wedge N^{\prime}(B,\varepsilon) \rightarrow \nstop$, it is mapped into a \textbf{return} statement guarded by a condition that is identical with that in line 3 (lines 9-10).
{\small
\begin{lstlisting}
// code for delayed continuous statement
for(int i=0;i<T/h;i++){
    if(N(B,e)&&N_p(B,e)){
        wait(h,SC_TU);
        x=x+h*f(x,x_r);
        wait(SC_ZERO_TIME);
    }
}
if(N(B,e)&&N_p(B,e)){
    return;
}
\end{lstlisting}
}

For space limitation, the rest of the code generation rules can  be found in Appendix B.
Thus now, for a given discretized {\dhcsp} process $\discrete{h}{\varepsilon}{S}$, we can generate its corresponding SystemC implementation $SC(\discrete{h}{\varepsilon}{S})$. Furthermore, their ``equivalence'' can be guaranteed by the following theorem.
\begin{theorem}
 For a {\dhcsp} process S, $\discrete{h}{\varepsilon}{S}$ and $SC(\discrete{h}{\varepsilon}{S})$ are bisimilar.
 \label{theorem:codegeneration}
\end{theorem}

%

\section{Case study}
\label{section:casestudy}
In this section, we illustrate how to generate SystemC code from {\dhcsp} through the example of water tank in
Exmaple~\ref{eg:wts-1}. 
As discussed above, for a given {\dhcsp} process, the procedure of code generation is divided into two steps: (1) compute the value of step size $h$ that can ensure the original {\dhcsp} process and its discretization are approximately bisimilar
with respect to the given precisions; (2) generate SystemC code from the discretized {\dhcsp} process.
We have implemented a tool that can generate code from both HCSP and {\dhcsp} processes \footnote{The tool and all examples for HCSP and {\dhcsp} can be found at \url{https://github.com/HCSP-CodeGeneration/HCSP2SystemC}.}.

Continue to consider 
Exmaple~\ref{eg:wts-1}. 
For given $h$, $\varepsilon$ and $T$, by using the discretized rules, a discretization system $\textit{WTS}_{h,\varepsilon}$ is obtained as follows:
{\small
\begin{eqnarray*}
\textit{WTS}_{h,\varepsilon} & \Define & \textit{Watertank}_{h,\varepsilon} \| \textit{Controller}_{h,\varepsilon} \\
\textit{Watertank}_{h,\varepsilon} & \Define & v:=v_0; d:=d_0;( v=1 \to (wl!:=1;(wl! \land \lnot wl? \to\\ & &
 (wait \ h;d(t+h)=d(t)+h(Q_{max} - \pi s^2  \sqrt{g (d(t)+d(t-r))}))^{\frac{T}{h}};\\ & &
 wl! \land wl? \to (wl!d;wl!:=0;cv?:=1;cv?v;\\ & & cv?:=0);
wl! \land \lnot wl? \to\nstop);\\ & &
v=0 \to (wl!:=1;
 (wl! \land \lnot wl? \to (wait \ h;\\ & & d(t+h)=d(t)+h(- \pi s^2  \sqrt{g (d(t)+d(t-r))})))^{\frac{T}{h}};\\ & &
wl! \land wl? \to (wl!d;wl!:=0;cv?:=1;cv?v;\\ & & cv?:=0);
wl! \land \lnot wl? \to\nstop)
)^* \\
\textit{Controller}_{h,\varepsilon} &\Define & y:=v_0;x:=d_0;(wait \ p; wl?:=1;wl?x;\\
 & & wl?:=0;  x\ge ub \to y:=0;x \le lb \to y:=1;\\
 & & cv!:=1;cv!y;cv!:=0)^*
\end{eqnarray*}
}

Given $Q_{max}=2.0$, $\pi=3.14$, $s=0.18$, $g=9.8$, $p=1$, $r=0.1$, $\textit{lb}=4.1$, $\textit{ub}=5.9$, $v_0=1$ and $d_0=4.5$,
we first build an instance of \emph{WTS} (the Watertank\_delay.hcsp file).
Then, according to the simulation result, we can estimate that the valid scope of $\delta$ and $\epsilon$ for \emph{WTS} is $\delta=0$ and $\epsilon \le 0.217$, respectively. By Theorem ~\ref{theorem:app-process}, we can infer that a discretized time step $h$ must exist s.t. \emph{WTS} and $\textit{WTS}_{h,\varepsilon}$ are $(h,\varepsilon)$-approximately bisimilar, with $\varepsilon \le \epsilon$.
For given values of $\varepsilon$ and time bound $T$, e.g., $\varepsilon = 0.2$ and $T=10$, we obtain $h=0.025$ (by Alg. \ref{alg:compt-h-multi-dde} in Sec. \ref{section:Dis ConD}) s.t. Theorem ~\ref{theorem:app-process} holds, i.e., $\textit{WTS}\cong_{h,\varepsilon}\discrete{h}{\varepsilon}{\textit{WTS}}$ on $[0,10]$.
After that, we can automatically generate  SystemC code equivalent to $\discrete{h}{\varepsilon}{\textit{WTS}}$ (by calling HCSP2SystemC.jar).

\begin{figure}[!htbp]
\centering
\subfloat[]{\includegraphics[scale=0.52]{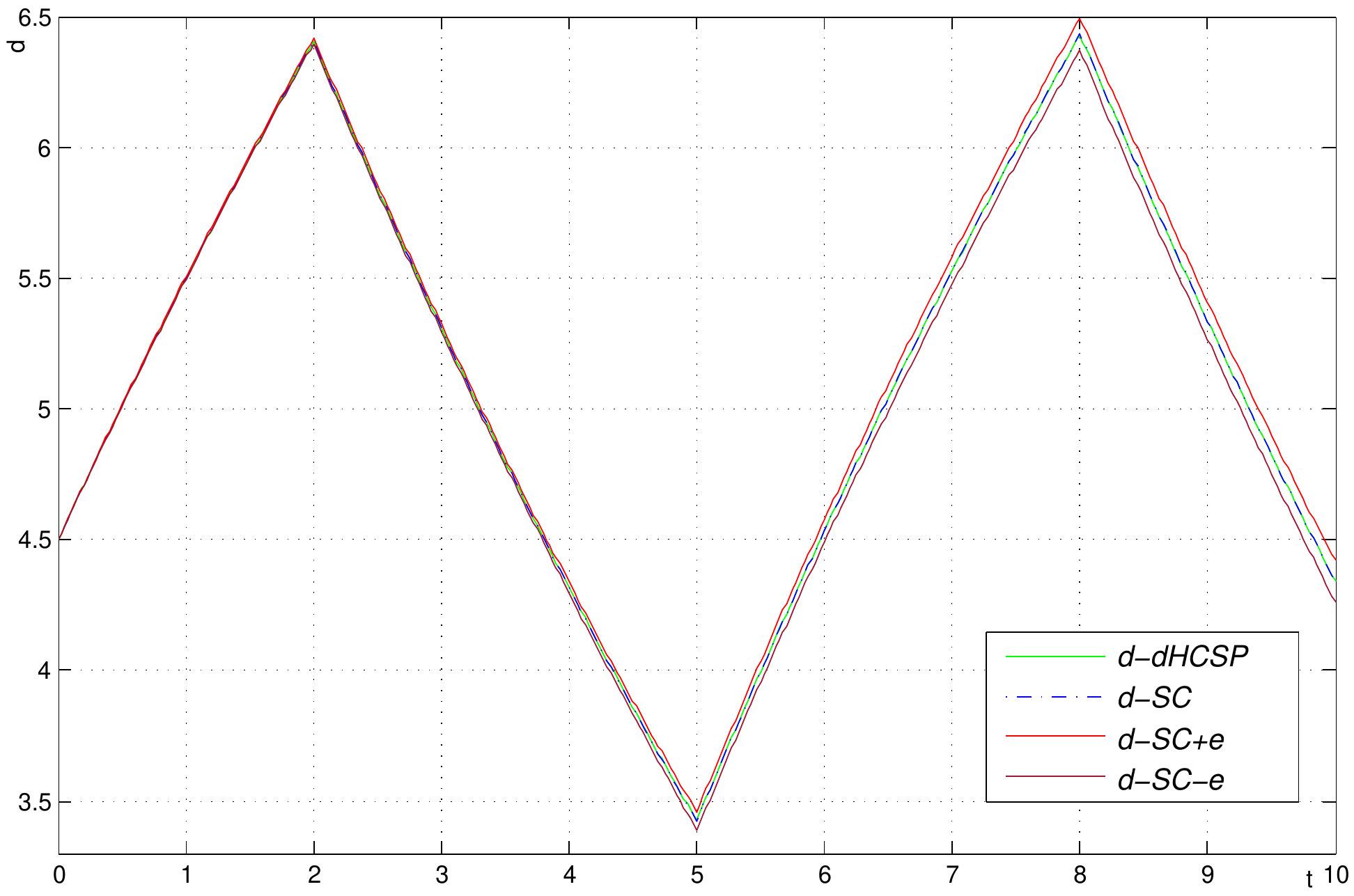}}\\
\subfloat[]{\includegraphics[scale=0.39]{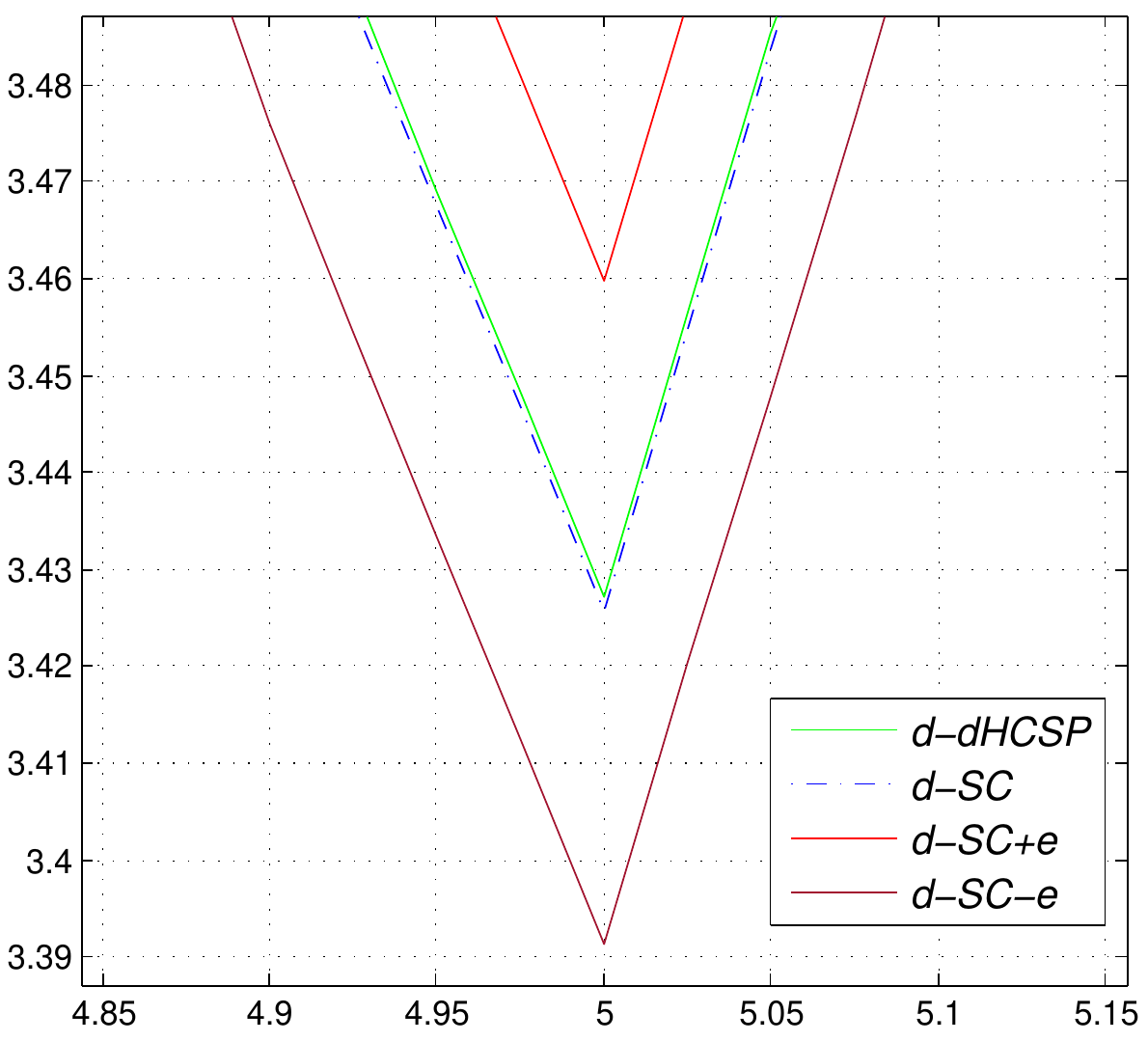}}\hspace{15pt}
\subfloat[]{\includegraphics[scale=0.39]{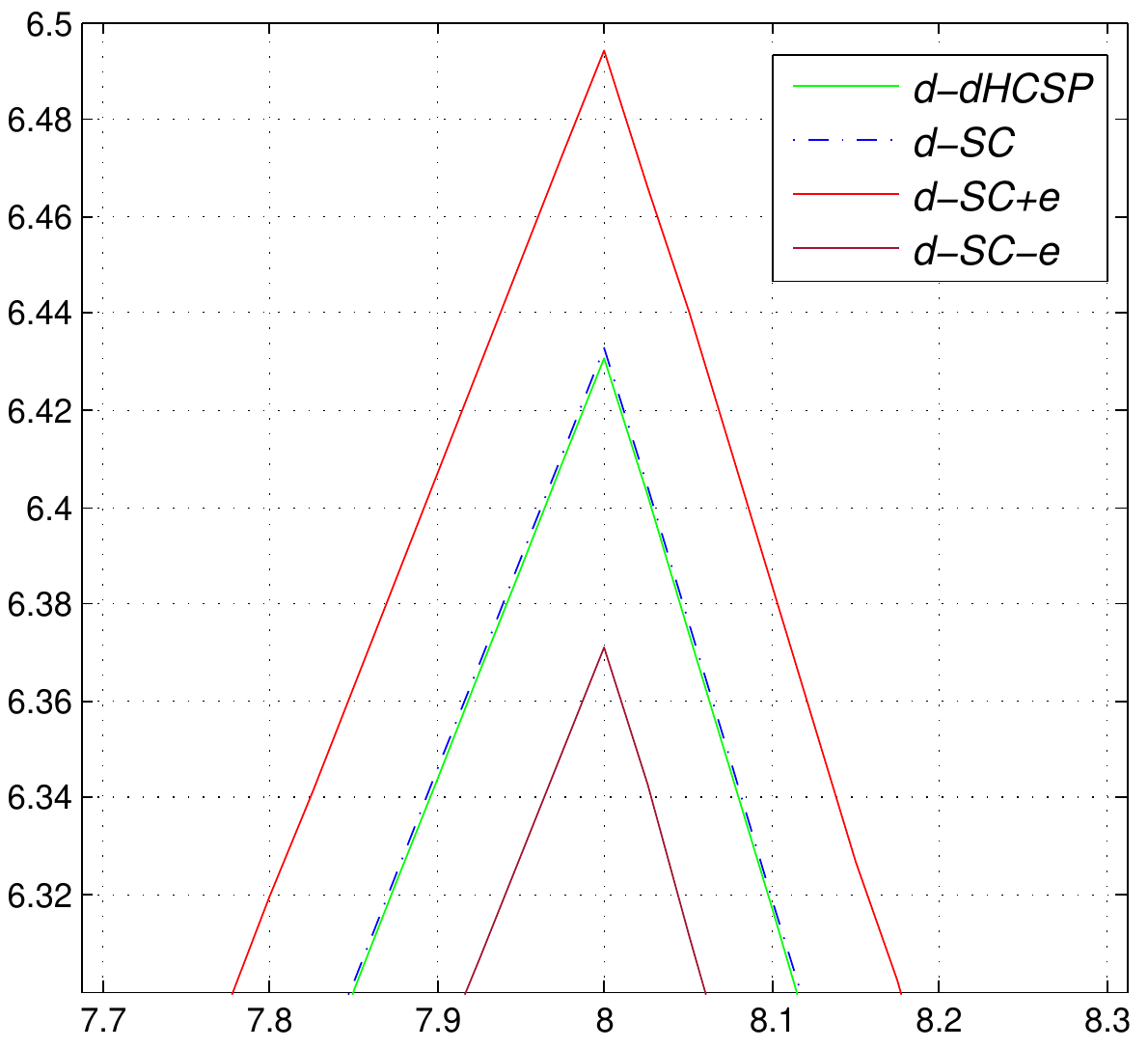}}
\caption{The {\dhcsp} model \emph{vs.} the SystemC code of $\textit{WTS}$ (Color figure online).(a) The result on [0,10]; (b) Zoom in on the result around 5; (c) Zoom in on the result around 8.}
\label{fig:WT-d}
\end{figure}

The comparison of the results, i.e., the curves of the water level ($d$ in the figure), which are acquired from the simulation of the original {\dhcsp} model and the generated SystemC code respectively is shown in Fig.~\ref{fig:WT-d}.
The result on the whole time interval $[0,10]$ is illustrated in Fig.~\ref{fig:WT-d} (a), and the specific details around two vital points, i.e., $5$ and $8$, are shown in Fig.~\ref{fig:WT-d} (b) and Fig.~\ref{fig:WT-d} (c), respectively.
In the figures, the simulation result (by calling the DDE solver $dde23$ in Matlab) is represented by green solid (i.e., \emph{d-{\dhcsp}}), and the result obtained by running the generated SystemC code is represented by blue dashed (i.e., \emph{d-SC}). The upper bound (lower bound) of the SystemC result, by adding (subtracting) the local error bounds computed in Alg. \ref{alg:compt-h-multi-dde}, is represented by red solid (dark red solid), i.e., \emph{d-SC+e} (\emph{d-SC-e}).
As Fig.~\ref{fig:WT-d} shows, the results of simulation and SystemC code both always fall into the interval determined by the upper and lower error bounds, which indicates the correctness of the discretization. Moreover, the distance between the state of the simulation and the state of SystemC code is less than the required precision (i.e., $\varepsilon = 0.2$), in every interval of $h$ length.

\section{Conclusion}
\label{section:conclusion}


In this paper, we present an automatic translation from abstract {\dhcsp} models to executable  SystemC code, while preserving approximate equivalence between them within given precisions. As a modelling language for hybrid systems, {\dhcsp} includes continuous dynamics in the form of DDEs and ODEs, discrete dynamics, and interactions between them based on communication, parallel composition and so on. In the discretization of {\dhcsp} within bounded time, on one hand, based on our previous work, we discretize a DDE by a sequence of approximate discrete states and control the distance from the trajectory within a given precision, by choosing a proper discretized time step to make the error bound less than the precision; and on the other hand, by requiring the original {\dhcsp} models to be robustly safe, we guarantee the consistency between the execution flows of the source model and its discretization in the sense of approximate bisimulation with respect to 
the given error tolerance. 


As a future work, we will continue to transform from SystemC code into other practical programming languages, such as C, C++, java, etc.  In addition, we also consider to apply our approach to more complicated real-world case studies.

\bibliographystyle{splncs03}
\bibliography{sigproc}

\begin{thebibliography}{10}
\providecommand{\url}[1]{\texttt{#1}}
\providecommand{\urlprefix}{URL }

\bibitem{RationalRose}
Rational Rose, \url{http://www-03.ibm.com/software/products/en/rosemod}

\bibitem{slusing}
Simulink, \url{https://cn.mathworks.com/products/simulink.html}

\bibitem{TargetLink}
TargetLink,
  \url{https://www.dspace.com/en/inc/home/products/sw/pcgs/targetli.cfm}

\bibitem{Ahmad14}
Ahmad, E., Dong, Y., Wang, S., Zhan, N., Zou, L.: Adding formal meanings to
  {AADL} with hybrid annex. In: FACS, pp. 228--247. Springer (2014)

\bibitem{CHARON}
Alur, R., Grosu, R., Hur, Y., Kumar, V., Lee, I.: Modular specification of
  hybrid systems in {CHARON}. In: {HSCC} 2000. pp. 6--19 (2000)

\bibitem{singlethread}
Alur, R., Ivancic, F., Kim, J., Lee, I., Sokolsky, O.: Generating embedded
  software from hierarchical hybrid models. In: {LCTES} 2003. pp. 171--182
  (2003)

\bibitem{TOC2010}
Anand, M., Fischmeister, S., Hur, Y., Kim, J., Lee, I.: Generating reliable
  code from hybrid-systems models. {IEEE} Trans. Computers  59(9),  1281--1294
  (2010)

\bibitem{TAC02Angeli}
Angeli, D., et~al.: A {Lyapunov} approach to incremental stability properties.
  IEEE Transactions on Automatic Control  47(3),  410--421 (2002)

\bibitem{Bellen2013}
Bellen, A., Zennaro, M.: Numerical methods for delay differential equations.
  Oxford university press (2013)

\bibitem{Esterel}
Berry, G.: The foundations of esterel. In: Proof, Language, and Interaction,
  Essays in Honour of Robin Milner. pp. 425--454 (2000)

\bibitem{Chen16}
{Chen}, M., {Fr{\"{a}}nzle}, M., {Li}, Y., {Mosaad}, P., {Zhan}, N.: Validated
  simulation-based verification of delayed differential dynamics. In: {FM}
  2016, LNCS, vol. 9995, pp. 137--154 (2016)

\bibitem{SHIFT}
Deshpande, A., G{\"{o}}ll{\"{u}}, A., Varaiya, P.: {SHIFT:} {A} formalism and a
  programming language for dynamic networks of hybrid automata. In: Hybrid
  Systems {IV}. pp. 113--133 (1996)

\bibitem{FraenzleEA:iSAT:JSAT}
Fr\"anzle, M., Herde, C., Ratschan, S., Schubert, T., Teige, T.: Efficient
  solving of large non-linear arithmetic constraint systems with complex
  {B}oolean structure. Journal on Satisfiability, Boolean Modeling and
  Computation  1,  209--236 (2007)

\bibitem{Girard07a}
Girard, A., Pappas, G.: Approximation metrics for discrete and continuous
  systems. IEEE {T}ransactions on {A}utomatic {C}ontrol  52(5),  782--798
  (2007)

\bibitem{Lustre1}
Halbwachs, N., Caspi, P., Raymond, P., Pilaud, D.: The synchronous dataflow
  programming language lustre. In: Proceedings of the IEEE. pp. 1305--1320
  (1991)

\bibitem{Statecharts}
Harel, D.: Statecharts: {A} visual formalism for complex systems. Sci. Comput.
  Program.  8(3),  231--274 (1987)

\bibitem{HS06}
Henzinger, T., Sifakis, J.: The embedded systems design challenge. In: FM 2006,
  LNCS, vol. 4085, pp. 1--15 (2006)

\bibitem{Huang17}
Huang, Z., Fan, C., Mitra, S.: Bounded invariant verification for time-delayed
  nonlinear networked dynamical systems. Nonlinear Analysis: Hybrid Systems
  23,  211--229 (2017)

\bibitem{multithread}
Hur, Y., Kim, J., Lee, I., Choi, J.: Sound code generation from communicating
  hybrid models. In: {HSCC} 2004. pp. 432--447 (2004)

\bibitem{Lee00}
Lee, E.: What's ahead for embedded software? Computer  33(9),  18--26 (2000)

\bibitem{Pola15}
Pola, G., Pepe, P., Di~Benedetto, M.: Symbolic models for nonlinear
  time-varying time-delay systems via alternating approximate bisimulation.
  Int. J. Robust Nonlinear Control  25(14),  2328--2347 (2015)

\bibitem{Pola10}
Pola, G., Pepe, P., Di~Benedetto, M., Tabuada, P.: Symbolic models for
  nonlinear time-delay systems using approximate bisimulations. Systems {\&}
  Control Letters  59(6),  365--373 (2010)

\bibitem{PJ05}
Prajna, S., Jadbabaie, A.: Methods for safety verification of time-delay
  systems. In: {CDC} 2005. pp. 4348--4353 (2005)

\bibitem{Yan16}
{Yan}, G., {Jiao}, L., {Li}, Y., {Wang}, S., {Zhan}, N.: Approximate
  bisimulation and discretization of hybrid {CSP}. In: {FM} 2016, LNCS, vol.
  9995, pp. 702--720 (2016)

\bibitem{Yan17}
Yan, G., Jiao, L., Wang, L., Wang, S., Zhan, N.: Automatically generating
  {SystemC} code from {HCSP} formal models  (Submitted)

\bibitem{Zhan16}
Zhan, N., Wang, S., Zhao, H.: Formal verification of Simulink/Stateflow
  diagrams: a deductive way. Springer (2016)

\bibitem{Zou15}
Zou, L., Fr{\"{a}}nzle, M., Zhan, N., Mosaad, P.: Automatic verification of
  stability and safety for delay differential equations. In: {CAV} 2015. LNCS,
  vol. 9207, pp. 338--355 (2015)

\end{thebibliography}

\newpage
\section{Appendix A}
\subsection{Proof of Theorem~\ref{theorem:decidable}}
\begin{proof}
In order to prove two {\dhcsp} processes, $S_1$ and $S_2$, are $(h, \varepsilon)$-approximately bisimilar on $[0,T]$ for given positive $h$, $\varepsilon$ and $T$, we need to find an $(h, \varepsilon)$-approximate bisimulation relation $\mathcal{B}_{h,\varepsilon}$  s.t. $(S_1, S_2) \in \mathcal{B}_{h,\varepsilon}$ on $[0,T]$ (Def. \ref{definition:TSappbisimulation}).
From the definition of the operational semantics of {\dhcsp}, we can construct a transition system from a {\dhcsp} process for a discrete time step size $d$. Within the acquired transition system, states are denoted as a set of pairs $(S_i,\staten_i)$, where $S_i$ is the remaining {\dhcsp} process will be executed and $\staten_i$ is the state of the {\dhcsp} process defined in \ref{section:semanticsofdhcsp}, and labels on transitions are identical with that in the {\dhcsp} process. For a {\dhcsp} process $S$, the transition system (denoted as $TS(S)$) constructed from it is \emph{symbolic} (containing finite states), since $T$ is bounded and constructors in $S$ is finite.

For two {\dhcsp} processes $S_1$ and $S_2$, we first construct their transition systems, i.e., $TS(S_1)$ and $TS(S_2)$ respectively, then we can compute a maximal approximate bisimulation
relation (satisfy the conditions in Def. \ref{definition:TSappbisimulation}) between $TS(S_1)$ and $TS(S_2)$ for the given step precision $h$ and state precision
$\varepsilon$, inspired by Algorithm 3 in \cite{Girard07a}.
After that, we can decide whether $TS(S_1)$ and $TS(S_2)$ are $(h,\varepsilon)$-approximately bisimilar,
depending on the fact whether all the initial states of $TS(S_1)$ and $TS(S_2)$ (for the {\dhcsp} process, $S_1$ and $S_2$ respectively) belong to the maximal approximate bisimulation
relation. As a result, if $TS(S_1)$ and $TS(S_2)$ are $(h,\varepsilon)$-approximately bisimilar on $[0,T]$, we can conclude that $S_1$ and $S_2$ are $(h,\varepsilon)$-approximately bisimilar on $[0,T]$. Since $TS(S_1)$ and $TS(S_2)$ are both \emph{symbolic}, the procedure for computing
the maximal approximate bisimulation relation can always terminate.

From the above illustration, we can conclude that the procedure for deciding whether
two {\dhcsp} processes are approximately bisimilar in bounded time is guaranteed to
terminate within finite time. Thus it is decidable.
\qed
\end{proof}

\subsection{Proof of Theorem~\ref{theorem:DDEDiscrete}}
\begin{proof}In general, we assume $T_d$ be an integral multiple of $h$. This assumption is reasonable, because we can always choose a $T_d^{\prime}\ge T_d$ s.t. $T_d^{\prime}$ is an integral multiple of $r$, and of course of $h$, to make the the DDE and its discretization are approximately bisimilar on $[0,T_d^{\prime}]$, so as on $[0,T_d]$.
For convenience sake, $\Gamma$ and $D(\Gamma)$ are used to denote the DDE and its discretization, respectively.

From Def.~\ref{definition:TSappbisimulation}, in order to prove that $\Gamma$ and $D(\Gamma)$ are $(h,\euler)$-approximately bisimilar, we
need to prove that there exists an
$(h,\euler)$-approximate bisimulation relation, $\mathcal{B}_{h,\euler}$,
between $\Gamma$ and $D(\Gamma)$ such that $(\Gamma,D(\Gamma)) \in \mathcal{B}_{h,\euler}$.
For the initial state $\rho_0$ (i.e., $\now$=0), $\|\xx_0 - \fg(0)\| =d_0<\euler$ holds obviously.
In order to illustrate the existence of $\mathcal{B}_{h,\euler}$, according to
Def.~\ref{definition:appbisimulation}, we should ensure that the
``distance'' between $\Gamma$ and $D(\Gamma)$ is never greater than $\euler$ within
all intervals $[t_i,t_{i+1}]$
with $i \in [0,n-1]$ (here $t_{i+1}-t_i=h$ and $n=\frac{T_d}{h}$), which can be guaranteed when the diameter of every $conv(N(\xx_i,d_i)\cup N(\xx_{i+1},d_{i+1}))$ is not greater than the precision $\euler$, i.e., $dia(conv(N(\xx_i,d_i)\cup N(\xx_{i+1},d_{i+1})))<\euler$ for all $i\in[0,n-1]$. As illustrated in \cite{Chen16}, we can always find small enough $h$ to make this constraint satisfied. In other words, for a given precision $\euler$, and a initial error $\|\xx_0 - \fg(0)\| <\euler$, we can always find a step size $h$ s.t. $\mathcal{B}_{h,\euler}$ exists and $(\Gamma,D(\Gamma)) \in \mathcal{B}_{h,\euler}$, so the theorem holds.
\qed
\end{proof}

\subsection{Proof of Theorem~\ref{theorem:app-process}}
\begin{proof}
For a {\dhcsp} process $S$, a given step size $h$ and time bound $T$, we prove that the global discretized
error between $S$ and $\discrete{h}{\varepsilon}{S}$ on $[0,T]$ (i.e., the maximal error for every $h$-length interval) is $Dh$,
for some constant $D$. As a result, when $h$ is sufficiently small (i.e., $h <\frac{\vare}{D}$),
$Dh < \vare$ is guaranteed. Then, with $S$ and $\discrete{h}{\varepsilon}{S}$ starting
execution from the same initial state $\rho_0$, we can conclude that
$S\cong_{h,\varepsilon}\discrete{h}{\varepsilon}{S}$ on $[0,T]$.

As $S$ and $\discrete{h}{\varepsilon}{S}$ start to execute from the same
initial state $\rho_0$, we suppose $S$ executes to $P$ with state $\sigma_1$,
and in correspondence, $\discrete{h}{\varepsilon}{S}$ executes to
$\discrete{h}{\varepsilon}{P}$ with some state $\beta_1$.
Denoting $\dd (\sigma_1, \beta_1)$ by  $\vare_1=D_1h$ for some $D_1$, and supposing $\vare_1 < \vare$, we prove that with $\vare_1$ as the initial error,
after the execution of $P$ and $\discrete{h}{\varepsilon}{P}$,
the global error (denoted by $\vare_2$) is $D_2 h$ for some constant $D_2$.
As a consequence, there must exist $h$ sufficiently small such that the global
error of $S$ is less than $\vare$. Notice that for the special case where $P$ is
$S$, $\vare_1$ is 0, and the above fact implies the theorem. Moreover, for the
satisfaction of $(\delta, \epsilon)$-robustly safe condition, two cases should be
considered here, i.e., $\delta=0$ and $\delta>0$.


For the first case that $\delta=0$, i.e., all boolean conditions in DDEs are \emph{true},
the DDEs may only be interrupted by the communication actions. In this case,
the DDE and its discretization have approximate control
flows and the difference between their ``jump'' time can be bounded by $h$.
The reason is that the execution time for any communication must fall into some $h$-length
duration and it can be detected within
$h$ in the discretized process.
Therefore, from the above description, we can always find $h$ sufficiently small to satisfy the
global discretized error constraint, such that $S\cong_{h,\varepsilon}\discrete{h}{\varepsilon}{S}$ on $[0,T]$.

For the second case that $\delta>0$, i.e, some DDEs whose boolean condition is not always $true$,
the DDEs may be interrupted by the violation of their boolean conditions. In this case, in order to
ensure that the DDE and its discretization have the approximate control
flow and the difference between their ``jump'' time can be bounded by $h$, an additional constraint
on $h$ (i.e., $\frac{\delta}{2}<h<\delta$), inferred from the $(\delta, \epsilon)$-robustly safe condition,
should also be satisfied. As a result, if there exists $\frac{\delta}{2}<h<\delta$ satisfying the
global discretized error constraint, we can conclude that $S\cong_{h,\varepsilon}\discrete{h}{\varepsilon}{S}$ on $[0,T]$.
Otherwise, we can not find a step size $h$ such that $S$ and $\discrete{h}{\varepsilon}{S}$ are $(h, \varepsilon)$-
approximately bisimilar on $[0,T]$.

For both cases, we first prove the existence of $h$ such that the global discretized
error constraint is satisfied, without considering the value of $\delta$. After that, if $\delta=0$ holds,
we can conclude that the scope of $h$ that we computed in the first step could make sure
$S\cong_{h,\varepsilon}\discrete{h}{\varepsilon}{S}$ on $[0,T]$. Otherwise, if $\delta>0$ and the scope
of $h$ got in the first step has overlaps with $(\frac{\delta}{2},\delta)$, we can also conclude
that the existence of $h$ such that $S\cong_{h,\varepsilon}\discrete{h}{\varepsilon}{S}$ on $[0,T]$.
However, if $\delta>0$ and the scope of $h$ does not fall into the interval $(\frac{\delta}{2},\delta)$,
we can conclude that there does not exist $h$ that makes $S$ and $\discrete{h}{\varepsilon}{S}$ be $(h, \varepsilon)$-
approximately bisimilar on $[0,T]$.

The proof of the the existence of $h$ such that the global discretized
error constraint is satisfied, i.e., after the execution of $P$ and $\discrete{h}{\varepsilon}{P}$
the global error is $D_2 h$ for some constant $D_2$, is given
by structural induction on $P$. Since rules of discretization for constructors in {\dhcsp} are closely similar to those in HCSP, except a slight difference for terms containing DDEs, so we only illustrate the proofs for these kinds of {\dhcsp} processes here, and proofs for other cases can refer to \cite{Yan16}.

\begin{itemize}
\oomit{\item Case $P_1=\pskip$: the discretized process is  $\pskip$. Obviously $\vare_2=\vare_1$.
   \item Case $P_1=(x:=e)$:  the discretized process  is  $x:=e$, where $e$ is an expression of variables, thus can be written as a function application of form $f(x_1, \cdots, x_n)$, among which $x_1, \cdots, x_n$ denote the variables occurring in $e$. After the assignment, only the value of $x$ is changed. Thus, from the definition of $\dd$, we have the fact $\vare_2 = \max(\vare_1, |a_2 - a_1|)$, in which $a_1 = \sigma_1(e)$ and $a_2 = \beta_1(e)$ represent the value of $x$ after the assignment. From the definition of $e$, $a_2 = f(\beta_1(x_1), \cdots, \beta_1(x_n))$. For each $i=1, \cdots, n$, there exists $\delta_i$ such that $\beta_1(x_i) = \sigma_1(x_i) + \delta_i$ and $|\delta_i| \leq \vare_1$. By the Lagrange Mean Value Theorem, the following equation holds:
       \[a_2= f(\sigma_1(x_1), \cdots, \sigma_1(x_n)) + \sum_{i=1}^n \frac{\partial f}{\partial x_i}(\sigma_1(x_i) + \theta \delta_i) \delta_i\]
       where $\theta \in (0, 1)$. From the fact $f(\sigma_1(x_1), \cdots, \sigma_1(x_n)) = a_1$,
       \[
       \begin{array}{ll}
       & |a_2 - a_1| =  |\sum_{i=1}^n \frac{\partial f}{\partial x_i}(\sigma_1(x_i) + \theta \delta_i) \delta_i| \\ & \leq \vare_1\sum_{i=1}^n\max_{o\in (\sigma_1(x_i) - \vare_1, \sigma_1(x_i) + \vare_1)}|\frac{\partial f}{\partial x_i}(o)|
       \end{array}
       \]
       $n$ is a constant, and $\frac{\partial f}{\partial x_i}$ is bounded in the interval $(\sigma_1(x_i) - \vare_1, \sigma_1(x_i) + \vare_1)$, thus $|a_2 - a_1|$ is bounded by a multiplication of $\vare_1$ with a bounded constant. $\vare_2$ is the maximum of $\vare_1$ and this upper bound of $|a_2 - a_1|$. The fact holds obviously.
   \item Case $P_1=\pwait\ d$:   the discretized process  is  $\pwait\ d$. Obviously $\vare_2 = \vare_1$.
   \item Case $P_1=ch?x$:  the discretized process is $ch?:=1; ch?x; ch?:=0$. Notice that the auxiliary readiness variable $ch?$ is added in the discretized process, however, it will not introduce errors. Thus, we only consider the error between the common variables, i.e. process variables, of $P_1$ and its discretization.  There are two cases for the transitions of $P_1$ and $\discrete{h}{\varepsilon}{P_1}$. The first case is waiting for some time units. For this case, if the waiting time is finite, then let  the time durations for both sides be the same, $\vare_2 = \vare_1$ holds obviously; if the waiting time is infinite, indicating that a deadlock occurs, $\vare_2 = \vare_1$ holds also. For the finite case, at some time,
        an event $ch?c$ occurs, where $c$ is the value received, and as a consequence, $x$ is assigned to $c$. For both sides, let the value received, denoted by $c_1$ and $c_2$ respectively, satisfy $|c_1-c_2| \leq \widehat{M}\vare_1$ for some constant $\widehat{M}$. As a result, after the performance of the events $ch?c_1$ and $ch?c_2$ respectively, $\vare_2 = \max\{\vare_1, \widehat{M}\vare_1\}$.
   \item Case $P_1=ch!e$: the discretized process is $ch!:=1; ch!e; ch!:=0$. Same to input, there are two cases for the transitions of $ch!e$. For the first case, let the time duration for both sides be the same, thus $\vare_2 = \vare_1$ obviously. For the second case, the events $ch!\sigma_1(e)$ and $ch!\beta_1(e)$ occur, and from the proof for assignment, there must exist a constant $\widehat{M}$ such that $|\beta_1(e) - \sigma_1(e)| < \widehat{M} \vare_1$ holds. No variable is changed as a consequence of an output, thus, after the communication, $\vare_2 = \vare_1$ still holds.

   \item Case $P_1 = Q; Q'$: the discretized process is $\discrete{h}{\varepsilon}{Q}; \discrete{h}{\varepsilon}{Q'}$. By induction hypothesis, assume the error after the execution of $Q$ with initial error $\vare_1$ is $\vare_m$, then $\vare_2=M_3 \vare_m$ and $\vare_m=M_4h$ for some constants $M_3, M_4$. $\vare_2 = M_3M_4 h$ holds.

       \item Case $P_1 = B \rightarrow Q$: the discretized process is $B\rightarrow \discrete{h}{\varepsilon}{Q}$. From the assumption $\dd (\sigma_1, \beta_1) = \vare_1=M_1h$, then there exists sufficiently small $h$ such that $\vare_1 < \vare$. Let $\vare < \epsilon$, then $\dd (\sigma_1, \beta_1) < \epsilon$.   $P$ is $(h, \epsilon)$-robustly safe, thus if $\sigma_1(B)$ is true, from the definition that the distance between $\sigma_1$ and any state that makes $\neg B$ true is larger than $\epsilon$, we can prove that $\beta_1(B)$ must be true. For this case, $Q$ and $\discrete{h}{\varepsilon}{Q}$ will be executed. By induction hypothesis, we have $\vare_2=M_3h$ for some constant $M_3$. Likewise, if $\sigma_1(B)$ is false, then $\beta_1(B)$ must be false. For this case, $P_1$ terminates immediately.  By induction hypothesis, $\vare_2=  \vare_1$, thus the fact holds.

       \item Case $P_1 = Q \sqcap Q'$: the discretized process is $\discrete{h}{\varepsilon}{Q} \sqcap \discrete{h}{\varepsilon}{Q'}$. There are two cases for the execution of both $P_1$ and its descretized process. By making the same choice, suppose $Q$ and $\discrete{h}{\varepsilon}{Q}$ are chosen to execute.
           By induction hypothesis, we have $\vare_2=M _3h$ for some constant $M_3$. The other case when $Q'$ and $\discrete{h}{\varepsilon}{Q'}$ are chosen can be proved similarly.
}
\item Case $P =  \evolutionn{\dot{\xx}=\ff(\xx,\xx_r)}{B}$: Let $X(t, \sigma_1(\xx))$
represent the trajectory of  $\dot{\xx}=\ff(\xx,\xx_r)$ with the initial value
$\sigma_1(\xx)$ at $t_0$. As we only care about the relation between $P$ and $\discrete{h}{\varepsilon}{P}$
on $[t_0,t_0+T]$, behaviors beyond $t_0+T$ are not taken into account. When $B=true$ (i.e., $\delta=0$),
the execution of $P$ is just like an ordinary DDE. According to
Theorem ~\ref{theorem:DDEDiscrete}, we can always find $h$ such that
$P$ (with $B=true$) and its discretization are $(h, \varepsilon)$-approximately bisimilar on $[t_0,t_0+T]$
with an initial error $\vare_1$.
When $B$ is not always $true$ (i.e., $\delta>0$), assume time starts from $0$
(i.e., $t_0=0$ for simplicity purpose) and is divided by $h$, which results in a
sequence $\{t_i\}$ with $t_{i+1}-t_i=h$ for all $i\in \mathbb{N}$. Suppose $B$ fails
to hold for some $X(t_f, \sigma_1(\xx))$ at time $t_f$ with
$t_f \in (t_N, t_{N+1})$ for some $N\in \mathbb{N}$.
Three cases should be considered. First, if $T\le t_{N}$, i.e., $B$ turns to be
$false$ after $T$, we can infer that for the execution before $T$, both
$N(B,\varepsilon)$ and $N^{\prime}(B,\varepsilon)$ are $true$. The reason is:
before $t_N$, $B$ keeps holding, so does $N(B,\varepsilon)$, then, from
$T\le t_{N}$ we know that only the value of $N(B,\varepsilon)$ at $t_{N+1}$
should be promised to be $true$ (as $N^{\prime}(B,\varepsilon)$ is the value
of $N(B,\varepsilon)$ at the next step), and it holds from the fact that
$\dd(X(t_{N}, \sigma_1(\xx)),\xx_{N+1})\le \varepsilon$
(guaranteed by Theorem ~\ref{theorem:DDEDiscrete}) and $B$ holds at $t_{N}$.
So, when $T\le t_{N}$, it is just the same as the situation in
Theorem ~\ref{theorem:DDEDiscrete}, and
$P\cong_{h,\varepsilon}\discrete{h}{\varepsilon}{P}$ on $[0,T]$ obviously.
Second, if $t_{N}<T< t_{N+1}$, similar to the $T\le t_N$ situation,
we should decide the value of $N(B,\varepsilon)$ at $t_{N+1}$, which
has been proved to be $true$. Therefore,
$N(B,\varepsilon)\wedge N^{\prime}(B,\varepsilon)=true$ at any time point
$t\in[0,T]$ and $P\cong_{h,\varepsilon}\discrete{h}{\varepsilon}{P}$ on
$[0,T]$. At last, if $T\ge t_{N+1}$, according to the definition of
$(\delta, \epsilon)$-robustly safe, $N(B,\varepsilon)$ will be $false$ at
$t_{N+2}$, so nothing will be done after $t_{N+1}$ for the discrete process
$\discrete{h}{\varepsilon}{P}$. According to the semantics of {\dhcsp},
the original process $P$ will also do nothing after $t_f$. Since
$t_f \in (t_N, t_{N+1})$, we have $|t_f-t_{N+1}|<h$.
By Theorem ~\ref{theorem:DDEDiscrete},
$\dd(X(t_{f}, \sigma_1(\xx)),\xx_{N+1})\le \varepsilon$ holds.
Therefore, $P\cong_{h,\varepsilon}\discrete{h}{\varepsilon}{P}$ also holds on $[0,T]$ when $T\ge t_{N+1}$.
Hence, for given $\varepsilon < \epsilon$ and $\delta>0$, as the $(\delta, \epsilon)$-robustly
safe condition requires $\frac{\delta}{2}<h<\delta$ and
Theorem ~\ref{theorem:DDEDiscrete} requires another constraint on $h$,
only when the scope of $h$ acquired from Theorem ~\ref{theorem:DDEDiscrete} has
overlaps with $(\frac{\delta}{2},\delta)$, we can ensure that
$P$ and $\discrete{h}{\varepsilon}{P}$ are $(h,\varepsilon)$-approximately
bisimilar on $[0,T]$. As a result, when there exists $h$ and $\varepsilon$ s.t. $P\cong_{h,\varepsilon}\discrete{h}{\varepsilon}{P}$ on $[0,T]$,
the global discretized error constraint is satisfied obviously, i.e., after the execution of $P$ and $\discrete{h}{\varepsilon}{P}$
the global error is $D_2 h$ for some constant $D_2$.

\item  Case $P = \exempt{\evolutionn{\dot{\xx}=\ff(\xx,\xx_r)}{B}}{i\in I}{io_i}{Q_i}$:
First of all, notice that in the discretization of $P$, the auxiliary variables
$io_i, \overline{io_i}$ are added for assisting the execution of interruption.
These variables do not introduce errors.  Let $X(t, \sigma_1(\xx))$ represent
the trajectory of  $\dot{\xx}=\ff(\xx,\xx_r)$ with initial value $\sigma_1(\xx)$.
In fact, the communication interrupt can be regarded as special boolean conditions.
Moreover, it has higher priority than ordinary boolean expressions.
Since it does not introduce errors, the proof is similar to that for the
continuous evolution, i.e., three cases should be considered according to the
``$false$'' time of the ``special'' boolean condition, i.e., interruption time.
\end{itemize}
Till now, we have proved that for every statement $P$ in $S$, the
global error is $D_2 h$ for some constant $D_2$. So we can infer that the global error of the discretization for $S$,
say $\vare_g$, is $Dh$ for some constant $D$. In order to satisfy the
$\vare$ precision constraint, $\vare_g < \vare$ should hold,
i.e., $h<\frac{\vare}{D}$. Furthermore, for $\delta=0$, we can always find an $h$ sufficiently
small to satisfy $h<\frac{\vare}{D}$ and then $S\cong_{h,\varepsilon}\discrete{h}{\varepsilon}{S}$ on $[0,T]$.
For $\delta>0$, only the satisfaction of $\frac{\vare}{D}>\frac{\delta}{2}$ can ensure the
existence of $h$ such that $S\cong_{h,\varepsilon}\discrete{h}{\varepsilon}{S}$ on $[0,T]$.
Otherwise, if $\frac{\vare}{D}>\frac{\delta}{2}$ does not hold
(i.e., $h$ does not fall into $(\frac{\delta}{2},\delta)$) with $\delta>0$,
there exists no $h$ such that $S\cong_{h,\varepsilon}\discrete{h}{\varepsilon}{S}$ on $[0,T]$.
The fact is thus proved.
\qed
\end{proof}

\subsection{Proof of Theorem~\ref{theorem:codegeneration}}
\begin{proof}
We prove that there exists a bisimulation relation $\mathcal{B}$, i.e. $h,\varepsilon$ are both 0 in Def. ~\ref{definition:appbisimulation}, between
$\discrete{h}{\varepsilon}{S}$ and $SC(\discrete{h}{\varepsilon}{S})$.
Now, suppose $\discrete{h}{\varepsilon}{S}$ executes up to $P$ with state
$p_1$, while at the same time, $SC(\discrete{h}{\varepsilon}{S})$ executes
up to $SC(P)$ with state $q_1$, and $(p_1,q_1)\in \mathcal{B}$.
According to Def.~\ref{definition:appbisimulation}, if $p_2$ is
reachable from $p_1$ by executing an action $l$, and there exists $q_2$
s.t it is reachable from $q_1$ by executing the same action $l$, and moreover
$(p_2,q_2)\in \mathcal{B}$ holds, then we can conclude
that $P$ and $SC(P)$ are bisimilar. Therefore, for $\discrete{h}{\varepsilon}{S}$
and $SC(\discrete{h}{\varepsilon}{S})$ starting from the same initial state
$\rho_0$, if all statements $P$ in $\discrete{h}{\varepsilon}{S}$ and the
corresponding $SC(P)$ in $SC(\discrete{h}{\varepsilon}{S})$ are bisimilar,
we can assure that $\discrete{h}{\varepsilon}{S}$ and
$SC(\discrete{h}{\varepsilon}{S})$ are bisimilar.
We can prove the bisimulation between $P$ and $SC(P)$ by
structural induction on $P$.
\begin{itemize}
\item Case P=($x:=e$): The execution of $x:=e$ is represented as a transition
$p_1 \xrightarrow{\tau} p_2$, in which $x$ equals to the value of the expression
$e$ in state $p_2$. Correspondingly, $q_1 \xrightarrow{\tau} q_2$ is generated
from the execution of $SC(P)$ (as \emph{wait(SC\_ZERO\_TIME)} changes nothing and
takes no time, we ignore its effect in the following), and the change of state
is identical with that in $P$. Since $(p_1,q_1)\in \mathcal{B}$ holds, we can
infer that $(p_2,q_2)\in \mathcal{B}$, thus $P$ and $SC(P)$ are bisimilar.
\item Case P=($wait \ d$): Since both $wait$ actions in $P$ and $SC(P)$ do not
change the state except for time advancing, we can easily conclude that for any
$0\le d^{\prime}\le d$ which makes $p_1 \xrightarrow{d^{\prime}} p_2$ happen in $P$,
there must exist a transition $q_1 \xrightarrow{d^{\prime}} q_2$ such that
$(p_2,q_2)\in \mathcal{B}$, and vice versa. Hence, $P$ and $SC(P)$ are bisimilar.
\item Case P=($B \rightarrow \discrete{h}{\varepsilon}{Q}$): For the
alternative statement, there are two cases. First, if $B$ is true at
$p_1$, it also holds at $q_1$ (as the distance between $p_1$ and $q_1$
is $0$). Assume $\discrete{h}{\varepsilon}{Q}$ and
$SC(\discrete{h}{\varepsilon}{Q})$ are bisimilar, $P$ and
$SC(P)$ are obviously bisimilar. Second, if $B$ does not hold at $p_1$
(neither at $q_1$), both the execution of $P$ and $SC(P)$ are represented
as a $\tau$ transition. So the bisimulation between $P$ and $SC(P)$
can also be promised.
\item Case P=($\discrete{h}{\varepsilon}{P_1}; \discrete{h}{\varepsilon}{P_2}$):
For the sequential composition,
suppose $\discrete{h}{\varepsilon}{P_1}$ and $SC(\discrete{h}{\varepsilon}{P_1})$
are bisimilar, then after the execution of them, the distance between $P$ and
$SC(P)$ is $0$. Moreover, if $\discrete{h}{\varepsilon}{P_2}$ and
$SC(\discrete{h}{\varepsilon}{P_2})$ are also bisimilar, the distance between
$P$ and $SC(P)$, after the execution of $\discrete{h}{\varepsilon}{P_2}$ and
$SC(\discrete{h}{\varepsilon}{P_2})$, can also be guaranteed to be $0$. Therefore,
$P$ and $SC(P)$ are bisimilar inductively.
\item Case P=($ch?:=1; ch?x;  ch?:=0$): In the SystemC implementation, we use
additional signals and events to ensure the synchronization of communication.
Although some extra statements are introduced, e.g., \emph{wait(ch\_w\_done)} and
\emph{ch\_r\_done.notify()} for the receiving side, the execution of them in fact
takes no time and does not influence the states before them. Therefore,
we can just regard them as \emph{wait(SC\_ZERO\_TIME)}, i.e., whose effect is ignored.
Starting from the initial state $p_1$, if $P$ takes a $\tau$ transition
$p_1 \xrightarrow{ch?:=1} p_2$, so can $SC(P)$
(i.e., $q_1 \xrightarrow{ch\_r=1} q_2$). Since $ch?$ and $ch\_r$ are identical
variables, there are no distance between $p_2$ and $q_2$. Then, executing from
$p_2$, there are two cases for $P$. First, it waits $d$ time units until the
finish of the write side. For this case, $SC(P)$ will also wait
$d$ time units from $q_2$ (the \emph{wait(ch\_w\_done)} statement).
Second, there is no waiting, i.e., P executes $ch?x$ directly. For this case,
$SC(P)$ will also execute the corresponding statement \emph{x=ch.read()}. Both
$ch?x$ and \emph{x=ch.read()} assign $x$ with the current value of the channel
$ch$. Hence for the both cases, the distance between the post states of $p_2$
and $q_2$ is $0$. At last, the execution of $ch?:=0$ in $P$ is also bisimular
with \emph{ch\_r =0} in $SC(P)$. In a word, all transitions in $P$ have corresponding
transitions in $SC(P)$, such that the distance between their source and target
states are both $0$, and vice versa. As
a result, $P$ and $SC(P)$ are bisimilar.
\item Case P=($ch!:=1; ch!e; ch!:=0$): The proof is similar to the case for
P=($ch?:=1; ch?x;  ch?:=0$).
\item Case P=($\forall i\in I. io_i :=1; \talloblong_{i\in I} io_i\rightarrow (\forall i\in I. io_i :=0; \discrete{h}{\varepsilon}{P_i})$):
Since the situation where choosing a channel from multiple ready ones
nondeterminately is extremely unusual in actual scenarios, we assume
that no more than one channel gets ready at the same moment. In the
SystemC part, we use arrays \emph{I}, \emph{IO} and \emph{IO\_d} to store the index
of channels, the readiness information of $io$s and their duals,
respectively. From the previous proof, the bisimulation between $P$ and
$SC(P)$ can be ensured by the guarantee that every sequential process
in $P$ and its corresponding description in $SC(P)$ are bisimilar. For
the first part (i.e., $\forall i\in I. io_i :=1$), both $P$ and $SC(P)$
set the readiness variables to be $1$ sequentially, so their bisimulation
can be proved by the scenario in which multiple assignments are executed sequentially.
For the second alternative part, it has four phases: (1) the process will wait
until another process which contains one of the dual action of $io_i$ gets ready
for communication; (2) then the corresponding communicate event will take place
and its index is recorded; (3) after the communication, all the readiness
information will be reset to $0$; (4) at last, the corresponding subsequent
process is executed. We now illustrate that the behaviors of $SC(P)$ are
identical with the four phases in $P$ respectively. In $SC(P)$, the waiting
phase is implemented with a \emph{wait} statement whose waiting event list
is the disjunction of the duals of all channels in \emph{I}. It stops waiting as soon as a
communication in \emph{I} gets ready. When a communication event is ready
(i.e., $IO[i]==1\&\&IO\_d[i]==1$), the corresponding sending or receiving
action will be taken (i.e., \emph{io\_i}), and its index is recorded in \emph{k},
then the loop ends. Afterwards, all readiness information is reset to
$0$ (i.e., $IO[i]=0$), and a following process \emph{SC(P[k])} is executed. So,
for the second part of $P$, its behaviors are identical with those in
$SC(P)$, and the bisimulation between them can be easily concluded. As a result,
$P$ and $SC(P)$ are bisimilar.
\item Case P is the discretized delayed continuous statement: The execution of $P$
consists of two sequential segments. From the sequential composition
property, we know that if each sequential segment in $P$ and its corresponding
SystemC code block in $SC(P)$ are bisimilar, $P$ and $SC(P)$ are bisimilar.
In the following, we show it can be satisfied for every segment. First, for
$(N(B,\varepsilon)\wedge N^{\prime}(B,\varepsilon) \rightarrow (\pwait \ h; \xx:=\xx+h \ff(\xx,\xx_r)))^{\frac{T}{h}}$,
it runs the alternative statement for $\frac{T}{h}$ times,
in which if the boolean condition holds, a wait and an assignment action are
sequentially executed, otherwise, nothing happens. It is easy to see that
the behaviors defined in the SystemC implementation are identical with the
above description. Therefore, their bisimulation can be inferred with ease.
Second, for the
$\nstop$ statement, if the boolean condition holds, a $\nstop$ action
will be taken and the state of $P$ will not vary ever, which means $P$ has
been running for $T$ time units and the behavior beyond $T$ will not be taken into
consideration. Otherwise, $P$ must have ended before $T$, and nothing
happens for the $\nstop$ statement. The second code block in $SC(P)$ has the
identical semantics: if the boolean expression is satisfied, the process
terminates soon (i.e., return), which means the statements following $SC(P)$
will not be executed. Otherwise, it continues to run. In a word, all the two
segments in $P$ and the corresponding ones in $SC(P)$ are bisimilar. Hence
the bisimulation between $P$ and $SC(P)$ holds.
\item Case P is a discretized delayed continuous statement with communication interrupt:
Since $P$ is the combination of a continuous and a communication statement,
the bisimulation between $P$ and $SC(P)$ can be inspired by the proofs for
these two kinds of statements.
\item Case P is compound statements: For the compound constructors
P=$\discrete{h}{\varepsilon}{P_1} \| \discrete{h}{\varepsilon}{P_2}$,
P=$\discrete{h}{\varepsilon}{P_1} \sqcap \discrete{h}{\varepsilon}{P_2}$
and P=$(\discrete{h}{\varepsilon}{P_1})^{\ast}$, the bisimulation between
$P$ and $SC(P)$ can be proved inductively.
\end{itemize}
Till now, we have proved that for all kinds of constructors in {\dhcsp}, its
discritized version and the SystemC code generated from it are bisimilar. Therefore, for any
{\dhcsp} model $S$, $\discrete{h}{\varepsilon}{S}$ is bisimilar to
$SC(\discrete{h}{\varepsilon}{S})$. The fact is thus proved.
\qed
\end{proof}

\section{Appendix B}
\subsection{Rules of code generation for other constructors}
\label{section:appendixB}
The SystemC code generated from
\[
ch!:=1; ch!e; ch!:=0;
\]
is:
\begin{lstlisting}
// code for output statement
ch_w=1;
wait(SC_ZERO_TIME);
if(!ch_r)
    wait(ch_r.posedge_event());
ch.write(e);
wait(SC_ZERO_TIME);
ch_w_done.notify();
wait(ch_r_done);
ch_w=0;
wait(SC_ZERO_TIME);
\end{lstlisting}

Like the input statement, additional \emph{sc\_signal} and \emph{sc\_event}, i.e., $ch\_w$ for the readiness of the channel and $ch\_w\_done$ for the done of the writing action respectively, are imported in order to ensure synchronization. The procedure described by the SystemC code is defined as follows: first, the signal $ch\_w$ is initialized as $1$, which means the channel $ch$ is ready for writing (lines 2-3); then, the writing process waits for the readiness information from the reading process (lines 4-5), i.e., $ch\_r=1$; then, the writing process writes the value of the expression $e$ into the channel (lines 6-7); afterwards, it informs the termination of its writing to other processes (line 8) and waits for the completeness of the reading side (line 9); at last, it resets the value of $ch\_w$ to $0$ (lines 10-11).

The SystemC code generated from the communication choice statement:
\[
\forall i\in I. io_i :=1; \talloblong_{i\in I} io_i\rightarrow (\forall i\in I. io_i :=0; \discrete{h}{\varepsilon}{Q_i})
\]
is:
\begin{lstlisting}
// code for communication choice statement
int k=-1;
int chan_num=sizeof(I)/sizeof(I[0]);
for(int i=0;i<chan_num;i++){
    IO[i]=1;
}
wait(SC_ZERO_TIME);
wait(IO_d[0].posedge_event()|...|
     IO_d[chan_num-1].posedge_event());
    for(int i=0;i<chan_num;i++){
        if(IO[i]==1&&IO_d[i]==1){
            io_i;
            k=i;
            break;
        }
    }
for(int i=0;i<chan_num;i++){
    IO[i]=0;
}
wait(SC_ZERO_TIME);
SC(Q[k]);
\end{lstlisting}
For simplicity, we assume the number of channels in the set $I$ is finite. Since the situation that more than one channel gets ready simultaneously is infrequent in the real system, we also assume that at most one channel gets ready in the same time. The implementation of the communication choice is illustrated as follows: first, the variable $k$, used for recording the index of the ready channel, is initialized as $-1$ and the number of channels in $I$ is recorded in an integer variable $chan\_num$ (lines 2-3); second, all the boolean variables corresponding to the channels in $I$ are set to $1$ (lines 4-7), i.e., the channels in $I$ are ready for reading or writing, which is the implementation of $\forall i\in I. io_i :=1$; third, like the single communication statement does, every reading (writing) process waits for the readiness of the writing (reading) side (lines 8-9); fourth, when the both sides of some channel get ready (lines 10-11), the communication happens (line 12) then its index is recorded in $k$ (line 13); after that, the boolean variables in $IO$ is reset to $0$ (lines 17-19); at last, the corresponding SystemC code of the process $Q[k]$ is executed (line 21).

The SystemC code generated from the communication interrupt statement:
\[  \begin{array}{c}
  \forall i\in I. io_i :=1;
  (N(B, \varepsilon)\wedge N^{\prime}(B,\varepsilon) \rightarrow \\
   \forall i\in I. io_i \wedge \neg \overline{io_i} \rightarrow
    (\pwait\ h; \xx:=\xx+h \ff(\xx,\xx_r)))^{\frac{T}{h}};\\
    \neg (N(B, \varepsilon)\wedge N^{\prime}(B,\varepsilon)) \wedge \forall i\in I. io_i \wedge \neg \overline{io_i} \rightarrow  \forall i\in I. io_i :=0;\\
     \exists i. io_i \wedge \overline{io_i} \rightarrow (\talloblong_{i\in I} io_i \rightarrow (\forall i\in I. io_i :=0; \discrete{h}{\varepsilon}{Q_i})); \\
    (N(B, \varepsilon)\wedge N^{\prime}(B,\varepsilon) \wedge \forall i\in I. io_i \wedge \neg \overline{io_i}) \rightarrow  \nstop;
   \end{array}
\]
is:
\begin{lstlisting}
// code for communication interrupt statement
int k=-1;
int chan_num=sizeof(I)/sizeof(I[0]);
for(int i=0;i<chan_num;i++){
    IO[i]=1;
}
wait(SC_ZERO_TIME);
for(int i=0;i<T/h;i++){
    if(N(B,e)&&N_p(B,e)&&IO[0]&&!IO_d[0]&&...){
        wait(h,SC_TU);
        x=x+h*f(x,x_r);
        wait(SC_ZERO_TIME);
    }
}
if(!(N(B,e)&&N_p(B,e))&&IO[0]&&!IO_d[0]&&...){
    for(int i=0;i<chan_num;i++){
        IO[i]=0;
    }
    wait(SC_ZERO_TIME);
}
for(int i=0;i<chan_num;i++){
    if(IO[i]==1&&IO_d[i]==1){
        io_i;
        k=i;
        break;
    }
}
for(int i=0;i<chan_num;i++){
    IO[i]=0;
}
wait(SC_ZERO_TIME);
if(k>-1){
    SC(Q[k]);
}
if(N(B,e)&&N_p(B,e)&&IO[0]&&!IO_d[0]&&...){
    return;
}
\end{lstlisting}

In order to generate the SystemC code from the discretized communication interrupt statement, we can combine the SystemC code generated from the discretized delayed continuous statement and the communication choice statement. Therefore, we have:
first, $\forall i\in I. io_i :=1$ is described by lines 2-7;
second, the discretized DDE is represented as a series of assignments and waits with the boolean condition (which is the combination of the neighborhood of $B$ in the delayed continuous statement and the readiness condition for the channels), for $\frac{T}{h}$ times of assignments with step size $h$ (lines 8-14);
third, for the situation that the neighborhood of the boolean condition is violated before some communication gets ready, the process terminates and all the boolean variables corresponding to the channels are reset to $0$ (lines 15-20);
afterwards, if some channel is ready before the boolean condition turns $false$, the relevant communication action happens and its index is recorded in $k$ (lines 21-27), then the boolean variables of the channels are reset to $0$ (lines 28-30) and the following code of the process $Q[k]$ is executed (lines 31-34);
at last, if the neighborhood of the boolean condition is never violated and no channels get ready during $[0,T]$, the whole process stops execution and returns (lines 35-37).

\end{document}